\newif\ifspringer
\springerfalse
\ifspringer
\documentclass{llncs}
\else
\documentclass{article}
\newtheorem{theorem}{Theorem}[section]
\newtheorem{proposition}[theorem]{Proposition}
\newtheorem{corollary}[theorem]{Corollary}
\newtheorem{lemma}[theorem]{Lemma}

\newenvironment{proof}{\emph{Proof.}}{\relax}

\newtheorem{example}[theorem]{Example}
\newcommand\qed{\hfill$\Box$\vskip0.2em}
\fi

\usepackage[latin1]{inputenc} 
\usepackage[T1]{fontenc}
\usepackage{graphicx}
\usepackage{amssymb}
\usepackage{amsfonts}
\usepackage{amsmath}
\usepackage{url}
\usepackage{stmaryrd}
\usepackage[all]{xypic}
\usepackage{latexsym}
\usepackage{prooftree}

\newcommand\nat{\mathbb{N}}
\newcommand\Z{\mathbb{Z}}
\newcommand\real{\mathbb{R}}

\newcommand\directed{\sideset{}{^{\,\makebox[0pt]{$\scriptstyle\uparrow$}\!}}}

\newcommand\dsup{\directed\sup}

\newcommand\dcup{\directed\bigcup}

\newcommand\Open{\mathcal O}
\newcommand\Eval[1]{\left\llbracket{#1}\right\rrbracket}

\newcommand\Rplus{\real_+}
\newcommand\creal{\overline\real_+}

\newcommand\cb[1]{\mathbf{#1}} 

\newcommand\identity[1]{\mathrm{id}_{#1}}
\newcommand\App{\text{\textsf{App}}}
\newcommand\lfp{\mathop{\mathrm{lfp}}\nolimits}
\newcommand{\Dcpo}{\cb{Dcpo}}
\newcommand\Dom{\mathop{\mathrm{dom}}}

\newcommand\eqdef{\mathrel{\buildrel\text{def}\over=}}
\newcommand\diff{\smallsetminus}

\newcommand\Val{\mathbf V}
\newcommand\Lform{{\mathcal L}}

\newcommand\ifkw{\mathop{\mathtt{if}}\nolimits}
\newcommand\elsekw{\mathbin{\mathtt{else}}}
\newcommand\ifc[3]{\ifkw {#1}\;\mathtt{then}\;{#2}\;\elsekw{#3}}
\newcommand\ifz[3]{\ifkw {#1}=0\;\mathtt{then}\;{#2}\;\elsekw{#3}}

\newcommand\letkw{\mathtt{let}}
\newcommand\letbe[2]{\letkw\;{#1}\;\mathtt{in}\;{#2}}
\newcommand\reckw{\mathop{\mathtt{rec}}\nolimits}
\newcommand\letreckw{\mathtt{letrec}}
\newcommand\letrecbe[2]{\letreckw\;{#1}\;\mathtt{in}\;{#2}}
\newcommand\recbe[1]{\reckw\;{#1}}
\newcommand\boolT{\mathtt{bool}}

\newcommand\intT{\mathtt{int}}

\newcommand\unitT{\mathtt{unit}}

\newcommand\suc{{\mathtt{s}}}
\newcommand\p{{\mathtt{p}}}

\newcommand\retkw{\mathop{\mathtt{ret}}\nolimits}
\newcommand\bindkw{\mathop{\mathtt{bind}}\nolimits}
\newcommand\dokw{\mathop{\mathtt{do}}}

\newcommand\step[1]{\mathrel{\buildrel {#1}\over\to}}
\newcommand\stepstar[1]{\mathrel{\buildrel {#1}\over\to{}^*}}

\newcommand\Prob{\mathop{\text{Pr}}}
\newcommand\vp{\mathbin{\downarrow}}
\newcommand\R{\mathrel{R}}
\newcommand{\sqsupsetsim}{\mathrel{{\raisebox{0.5ex}{$\mathop{\kern0pt \sqsupset}\limits_{\textstyle\sim}$}}}}
\newcommand{\sqsubsetsim}{\mathrel{{\raisebox{0.5ex}{$\mathop{\kern0pt \sqsubset}\limits_{\textstyle\sim}$}}}}

\newcommand\Nature{{\mathtt{P}}}

\newcommand\tr{{\triangleright}}

\title{Fooling the Parallel Or Tester with Probability $8/27$}

\newcommand\me{Jean Goubault-Larrecq}
\newcommand\lsv{LSV, 
  ENS Paris-Saclay, CNRS, Universit\'e Paris-Saclay, France\\
  Email:
  \email{goubault@lsv.fr}}
\newcommand\digicosme{%
  This research was partially supported by Labex
  DigiCosme (project ANR-11-LABEX-0045-DIGICOSME) operated by ANR as
  part of the program ``Investissement d'Avenir'' Idex Paris-Saclay
  (ANR-11-IDEX-0003-02).
}
\ifspringer
\author{\me\thanks{\digicosme}}
\institute{\lsv}
\else
\newcommand\email[1]{\url{#1}}
\author{\me\thanks{\digicosme} \\ \lsv}
\fi

\begin{document}


\maketitle

\begin{abstract}
  It is well-known that the higher-order language PCF is not fully
  abstract: there is a program---the so-called parallel or tester,
  meant to test whether its input behaves as a parallel or---which
  never terminates on any input, operationally, but is denotationally
  non-trivial.  We explore a probabilistic variant of PCF, and ask
  whether the parallel or tester exhibits a similar behavior there.
  The answer is no: operationally, one can feed the parallel or tester
  an input that will fool it into thinking it is a parallel or.  We
  show that the largest probability of success of such would-be
  parallel ors is exactly $8/27$.  The bound is reached by a very
  simple probabilistic program.  The difficult part is to show that
  that bound cannot be exceeded.
\end{abstract}

\ifspringer
\section{Capsule}

Catuscia and I have collaborated for a few years, but that was quite
some time ago \cite{GPT-aplas07}.  We were both working in computer
security then, and I have diverged since.  I still felt I had to
contribute to this Festschrift, but anything I could contribute in
computer security would be outdated, the few things I have worked on
in process algebra have more of an algebraic topological flavor, and
so the most relevant of Catuscia's research interests that I could
contribute to is probabilistic transition systems.  Here is a small
result related to that theme.  \fi

\section{Introduction}

There is a recurring theme in security: to defeat a strong adversary,
you need to rely on random choice.  This paper will be a somewhat
devious illustration of that principle, in the field of programming
language semantics.

The higher-order, functional language PCF \cite{Plotkin:PCF} forms the
core of actual programming languages such as Haskell
\cite{Bird:Haskell}.  Plotkin \cite{Plotkin:PCF}, and independently
Sazonov \cite{Sazonov:LCF}, had shown that PCF, while being adequate
(i.e., its operational and denotational semantics match, in a precise
sense), is not fully abstract: there are programs that are
contextually equivalent (a notion arising from the operational
semantics), but have different denotational semantics.  (One should
note that, conversely, two programs with the same denotational
semantics are always contextually equivalent.)

The argument is as follows.  In the denotational model, there is a
function of type $\intT \to \intT \to \intT$ called \emph{parallel
  or}, which maps the pair $1, 1$ to $1$, and both $0, N$ and $N, 0$
to $0$, for whatever program $N$ (including non-terminating programs).
One can show that parallel or is undefinable in PCF.  More is true.
One can define a PCF program, the \emph{parallel or tester}, which
takes an argument $f \colon \intT \to \intT \to \intT$, and tests
whether $f$ is a parallel or, by testing whether $f 1 1=1$,
$f 0 \Omega=0$, and $f \Omega 0=0$, where $\Omega$ is a canonical
non-terminating program.  The parallel or tester is contextually
equivalent to the always non-terminating program $\lambda f . \Omega$,
meaning that applying it to any PCF program (for $f$) will never
terminate.  However, the denotational semantics of the parallel or
tester and of $\lambda f . \Omega$ differ: applied to any given
parallel or map (which exists in the denotational model), one returns
and the other one does not.

We introduce a probabilistic variant of PCF which we call
PCF$_\Nature$, and we define a suitable parallel or tester
$\mathtt{portest}$.  A PCF$_\Nature$ program $M$ \emph{fools} the
parallel or tester if $\mathtt{portest}$ applied to $M$ terminates.
In PCF, there is no way of fooling the parallel or tester.  Our
purpose is to show that one can fool the parallel or tester of
PCF$_\Nature$ with probability at most $8/27$, and that this bound is
attained.  The optimal fooler is easy to define.  The hard part is to
show that one cannot do better.

A final word before we start.  Even though we started by motivating it
from matters related to full abstraction, which involves both
operational and denotational semantics, the question we are addressing
is purely \emph{operational} in nature: it is only concerned with the
behavior of $\mathtt{portest}$ under its operational semantics, under
arbitrary PCF$_\Nature$ contexts.  Nonetheless, denotational semantics
will be essential in our proof.

\emph{Outline.} We define the syntax of $PCF_\Nature$ in
Section~\ref{sec:syntax-pcf_nature}, its operational semantics in
Section~\ref{sec:oper-semant}, and---once we have stated the required
basic facts we need from domain theory in
Section~\ref{sec:refr-doma-theory}---its denotational semantics in
Section~\ref{sec:denot-semant}.  We state the adequacy theorem at the
end of the latter section.  This says that the operational and
denotational probabilities that a term $M$ of type $\intT$ terminates
on any given value $n \in \Z$ are the same.  We define the parallel
tester, and show that it can be fooled with probability $8/27$ at
most, in Section~\ref{sec:parallel-or-tester}.  We conclude by citing
some recent related work in Section~\ref{sec:conc}.



\section{The syntax of PCF$_\Nature$}
\label{sec:syntax-pcf_nature}

PCF$_\Nature$ is a typed language.  The \emph{types} are given by
the grammar:
\begin{align*}
  \sigma, \tau, \cdots
  & ::= \intT & \text{basic types} \\
  & \mid D \tau & \text{type of (subprobability) distributions on
                   $\tau$} \\
  & \mid \sigma \to \tau & \text{function types}.
\end{align*}
Mathematically, $D \tau$ will be the type of subprobability
valuations of elements of type $\tau$.  Operationally, an element of
type $D\tau$ is just a random value of type $\tau$.  There is only one
basic type, $\intT$, but one could envision a more expressive algebra
of datatypes.

A \emph{computation type} is a type of the form $D\tau$ or $\sigma
\to \tau$ where $\tau$ is a computation type.  The computation types
are the types where one can do computation, in particular whose objets
can be defined by recursion.

Our language will have functions, and a function mapping inputs of
type $\sigma$ to outputs of type $\tau$ will have type
$\sigma \to \tau$.
We write $\sigma_1 \to \sigma_2 \to \cdots \to \sigma_n \to \tau$ for
$\sigma_1 \to (\sigma_2 \to (\cdots \to (\sigma_n \to \tau) \cdots))$,
and this is a type of functions taking $n$ inputs, of respective types
$\sigma_1$, $\sigma_2$, \ldots, $\sigma_n$ and returning outputs of
type $\tau$.

We fix a countably infinite set of \emph{variables} $x_\tau$,
$y_\tau$, $z_\tau$, \ldots, for each type $\tau$.  Each variable has a
unique type, which we read off from its subscript.  We will
occasionally omit the type subscript when it is clear from context, or
irrelevant.

\begin{figure}
  \centering
  \[
  \begin{array}{c@{\qquad}c}
    \begin{prooftree}
      \justifies
      x_\tau \colon \tau
    \end{prooftree}
    \qquad
    \begin{prooftree}
      \justifies
      n \colon \intT
      \using (n \in \Z)
    \end{prooftree}
    &
    \begin{prooftree}
      M \colon \intT
      \justifies
      \suc M \colon \intT
    \end{prooftree}
    \qquad
      \begin{prooftree}
        M \colon \intT
        \justifies
        \p M \colon \intT        
      \end{prooftree}
    \\ \\
    \begin{prooftree}
      M \colon \intT \quad
      N \colon \tau \quad
      P \colon \tau
      \justifies
      \ifz M N P \colon \tau
    \end{prooftree}
    &
    \begin{prooftree}
      M \colon \tau \to \tau
      \justifies
      \reckw_\tau M
      \using (\tau \text{ computation type})
    \end{prooftree}
    \\ \\
    \begin{prooftree}
      M \colon \sigma \to \tau \quad
      N \colon \sigma
      \justifies
      MN \colon \tau
    \end{prooftree}
    &
      \begin{prooftree}
        M \colon \tau
        \justifies
        \lambda x_\sigma . M \colon \sigma \to \tau
      \end{prooftree}
    \\ \\
    \begin{prooftree}
      M \colon D\tau \quad
      N \colon D\tau
      \justifies
      M \oplus N \colon D \tau
    \end{prooftree}
    \qquad
    \begin{prooftree}
      M \colon \sigma
      \justifies
      \retkw_\sigma M \colon D\sigma
    \end{prooftree}
    &
      \begin{prooftree}
        M \colon D\sigma \quad
        N \colon \sigma \to D\tau
        \justifies
        \bindkw_{\sigma, \tau} M N \colon D\tau
      \end{prooftree}
  \end{array}
  \]
  \caption{The syntax of PCF$_\Nature$}
  \label{fig:syntax}
\end{figure}
The \emph{terms} $M$, $N$, \ldots, of our language are defined
inductively, together with their types, in Figure~\ref{fig:syntax}.
We agree to write $M \colon \tau$ to mean ``$M$ is a term, of type
$\tau$''.  We shall write $MN_1 N_2 \cdots N_n$ for
$(\cdots ((MN_1) N_2) \cdots) N_n$, and $\lambda x_1, \cdots, x_n . M$
for $\lambda x_1 . \lambda x_2 . \cdots . \lambda x_n . M$.  We shall
also use the abbreviations $\letbe{x_\sigma = M} N$ for
$(\lambda x_\sigma . N) M$ and $\letrecbe {f_\tau=M} N$, where
$M \colon \tau$, for
$\letbe {f_\tau = \recbe (\lambda f_\tau . M)} N$.  Finally, we shall
write $\dokw {x_\sigma \leftarrow M}; N$ for
$\bindkw_{\sigma,\tau} M (\lambda x_\sigma . N)$, of type $D \tau$
(draw $x_\sigma$ at random along distribution $M$, then run $N$).
$M \oplus N$ is meant to execute either $M$ or $N$ with probability
$1/2$.

The free variables and the bound variables of a term $M$ are defined
as usual.  A term with no free variable is \emph{ground}.  For a
substitution $\theta \eqdef [x_1:=N_1, \cdots, x_k:=N_k]$ (where each
$N_i$ has the same type as $x_i$, and the variables $x_i$ are pairwise
distinct), we write $M \theta$ for the parallel substitution of each
$N_i$ for each $x_i$, and $\Dom \theta$ for $\{x_1, \cdots, x_k\}$.
We say that $\theta$ is \emph{ground} if $N_1$, \ldots, $N_k$ are all
ground.

\begin{example}
  \label{ex:randnat}
  The term
  $\mathtt{rand\_int} \eqdef \reckw_{\intT \to D\intT} (\lambda
  r 
  . \lambda m_{\intT} . r (\suc m) \oplus \allowbreak \retkw_{\intT}
  m) 0$ is of type $D\intT$.  As we will see, this draws a natural
  number $n$ at random, with probability $1/2^{n+1}$.
\end{example}

\begin{example}
  \label{ex:rej:sample:3}
  \emph{Rejection sampling} is a process by which one draws an element
  of a subset $A$ of a space $X$, as follows: we draw an element of
  $X$ at random, and we return it if it lies in $A$, otherwise we
  start all over again.  Here is a simple example of rejection
  sampling, meant to draw a number uniformly among $\{0, 1, 2\}$.  The
  idea is to draw two independent bits at random, representing a
  number in $X \eqdef \{0, 1, 2, 3\}$, and to use rejection sampling
  on $A \eqdef \{0, 1, 2\}$.  Formally, we define the PCF$_\Nature$
  term
  $\mathtt{rand3} \eqdef \reckw_{D\intT} (\lambda p_{D\intT}
  . ((\retkw_\intT 0 \oplus \retkw_\intT 1) \oplus (\retkw_\intT 2
  \oplus p_{D\intT})))$.  Note that this uses recursion to define a
  distribution, not a function.
\end{example}

\section{Operational semantics}
\label{sec:oper-semant}

The \emph{elementary contexts} $E$, with their types
$\sigma \vdash \tau$, are defined as:
\begin{itemize}
\item $[\_ N]$ of type $(\sigma \to \tau) \vdash \tau$, for every $N
  \colon \sigma$, and for every type $\tau$;
\item $[\suc \_]$ and $[\p \_]$, of type $\intT \vdash \intT$;
\item $[\ifz \_ N P]$, of type $\intT \vdash \tau$, for all $N, P
  \colon \tau$;
\item $[\bindkw_{\sigma,\tau} \_ N]$, of type $D \sigma \vdash D
  \tau$, for every $N \colon \sigma \to D\tau$.
\end{itemize}
The \emph{initial contexts} are $[\_]$ (of type $\sigma \vdash \sigma$
for any $\sigma$) and $[\retkw_{\intT} \_]$ (of type
$\intT \vdash D \intT$).  The (evaluation) \emph{contexts} $C$ are the
finite sequences $E_0 E_1 \cdots E_n$, $n \in \nat$, where $E_0$ is an
initial context of type $\sigma_1 \vdash \sigma_0$, each $E_i$
($1\leq i \leq n$) is an elementary context of type
$\sigma_{i+1} \vdash \sigma_i$.  Then we say that $C$ has type
$\sigma_{n+1} \vdash \sigma_0$.

The notation $C [M]$ makes sense for every context
$C \eqdef E_0 E_1 \cdots E_n$ of type $\sigma \vdash \tau$ and every
$M \colon \sigma$, and is defined as $E_0 [E_1 [\cdots [E_n [M]]]]$,
where $E [M]$ is defined by removing the square brackets in $E$ and
replacing the hole $\_$ by $M$.  E.g., if
$C = [\retkw_{\intT} \_] [\p \_]$, then
$C [M] = \retkw_{\intT} (\p M)$.

\begin{figure}
  \centering
  \[
    \begin{array}{cc}
      \hline
      \multicolumn{2}{c}{\text{Exploration rules}} \\
      \hline
      C \cdot E [M] \step 1 C E \cdot M \quad \text{($E$ elem.\ context)}
      &
        [\_] \cdot \retkw_{\intT} M
        \step 1 [\retkw_{\intT} \_] \cdot M
      \\
      \hline
      \hline
      \multicolumn{2}{c}{\text{Computation rules}} \\
      \hline
      C [\_ N] \cdot \lambda x_\sigma . M 
     \step 1 C \cdot M [x_\sigma:=N]
      &  
        C \cdot \reckw_\tau M \step 1 C \cdot M (\reckw_\tau M)
      \\
        C \cdot M \oplus N \step {1/2} C \cdot M
      & 
        C \cdot M \oplus N \step {1/2} C \cdot N
      \\
        C [\bindkw_{\sigma,\tau} \_ N] \cdot \retkw_\sigma M 
      \step 1 C \cdot NM
      &
        C [\p \_] \cdot n \step 1 C \cdot n-1 \quad
        C [\suc \_] \cdot n \step 1 C \cdot n+1
      \\
      C [\ifz \_ N P] \cdot 0 \step 1 C \cdot N &
      C [\ifz \_ N P] \cdot n \step 1 C \cdot P \quad (n \neq 0)
      \\
      \hline
    \end{array}
  \]
  \caption{Operational semantics}
  \label{fig:opsem}
\end{figure}

A \emph{configuration} (of type $\tau$) is a pair $C \cdot M$, where
$C$ is a context of type $\sigma \vdash \tau$ and $M \colon \sigma$.

The operational semantics of PCF$_\Nature$---an abstract interpreter
that runs PCF$_\Nature$ programs---is a probabilistic transition
system on configurations, defined by the rules of
Figure~\ref{fig:opsem}.  We write $s \step \alpha s'$ to
say that one can go from configuration $s$ to configuration $s'$ in
one step, with probability $\alpha$.


A \emph{trace} is a sequence
$s_0 \step {\alpha_1} s_1 \step {\alpha_2} \cdots \step {\alpha_m}
s_m$, where $m \in \nat$, and where each
$s_{i-1} \step {\alpha_i} s_i$ is an instance of a rule of
Figure~\ref{fig:opsem}.  The trace starts at $s_0$, ends at $s_m$, its
\emph{length} is $m$ and its \emph{weight} is the product
$\alpha \eqdef \alpha_1 \cdots \alpha_2 \cdots \alpha_m$.  In that
case, we also write $s_0 \stepstar \alpha s_m$.

The \emph{run} starting at $s_0$ is the tree of all traces starting
at $s_0$.  Its root is $s_0$ itself, and for each vertex $s$ in the
tree, for each instance of a rule of the form $s \step \alpha t$, $t$
is a successor of $s$, and the edge from $s$ to $t$ is labeled
$\alpha$.

\begin{figure}
  \centering\footnotesize
    \[
    \xymatrix@R-18pt{
      [\_] \cdot \mathtt{rand\_int}
      \ar[d]^1
      \\
      [\_ 0] \cdot \reckw_{\intT \to D\intT} f
      \ar[d]^1
      & (f \eqdef \lambda r 
      . \lambda m 
      .  r (\suc m) \oplus \retkw_{\intT} m)
      \\
      [\_ 0] \cdot f (\reckw_{\intT \to D\intT} f)
      \ar[d]^1
      \\
      [\_ 0] [\_ (\reckw_{\intT \to D\intT} f)] \cdot \lambda
      r . \lambda m . r (\suc m) \oplus \retkw_{\intT} m
      \ar[d]^1
      \\
      [\_ 0] \cdot \lambda m .
      (\reckw_{\intT \to D\intT} f) (\suc m) \oplus \retkw_{\intT} m
      \ar[d]^1
      \\
      [\_] \cdot
      (\reckw_{\intT \to D\intT} f) (\suc 0) \oplus \retkw_{\intT} 0
      \ar[d]^{1/2} \ar[rd]^{1/2}
      \\
      [\_] \cdot
      (\reckw_{\intT \to D\intT} f) (\suc 0)
      \ar[dd]^1^(0.8){*}
      & [\_] \cdot \retkw_{\intT} 0 \ar[d]^1
      \\
      & [\retkw_{\intT} \_] \cdot 0
      \\
      [\_] \cdot
      (\reckw_{\intT \to D\intT} f) (\suc (\suc 0)) \oplus
      \retkw_{\intT} (\suc 0)
      \ar[d]^{1/2} \ar[rd]^{1/2}
      \\
      [\_] \cdot
      (\reckw_{\intT \to D\intT} f) (\suc (\suc 0))
      \ar[dd]^1^(0.8){*}
      & [\_] \cdot \retkw_{\intT} (\suc 0) \ar[d]^1^(0.7){*}
      \\
      & [\retkw_{\intT} \_] \cdot 1
      \\
      [\_] \cdot
      (\reckw_{\intT \to D\intT} f) (\suc (\suc (\suc 0))) \oplus
      \retkw_{\intT} (\suc (\suc 0))
      \ar[d]^{1/2} \ar[rd]^{1/2}
      \\
      [\_] \cdot
      (\reckw_{\intT \to D\intT} f) (\suc (\suc (\suc 0)))
      \ar[dd]^1^(0.7){*}
      & [\_] \cdot \retkw_{\intT} (\suc (\suc 0)) \ar[d]^1^(0.7){*}
      \\
      & [\retkw_{\intT} \_] \cdot 2 \\
      \vdots &
    }
  \]
  \caption{An example run in PCF$_\Nature$}
  \label{fig:randnat}
\end{figure}

For every configuration $s$ of type $D\intT$, and every $n \in \Z$, we
define $\Prob [s \vp n]$ as the sum of the weights of all traces that
start at $s$ and end at $[\retkw_{\intT} \_] \cdot n$.  This is the
\emph{subprobability that $s$ eventually computes $n$}.
We also write $\Prob [M \vp n]$ for $\Prob [[\_] \cdot M \vp n]$,
where $M \colon D\intT$.

\begin{example}
  \label{ex:randnat:op}
  The run starting at $\mathtt{rand\_int}$ (see
  Example~\ref{ex:randnat}) is shown in Figure~\ref{fig:randnat}.  We
  have abbreviated some sequences of $\step 1$ steps as $\stepstar 1$.
  One sees that $\Prob [\mathtt{rand\_int} \vp n] = 1/2^{n+1}$ for
  every $n \in \nat$, and is zero for every $n < 0$.  Notice the
  infinite branch on the left, whose weight is $0$.
\end{example}

\begin{example}
  \label{ex:rand3:op}
  We let the reader draw the run starting at $\mathtt{rand3}$ (see
  Example~\ref{ex:rej:sample:3}), and check that
  $\Prob [\mathtt{rand3} \vp n]$ is equal to $1/3$ if
  $n \in \{0, 1, 2\}$, $0$ otherwise.  Explicitly, if
  $n \in \{0, 1, 2\}$, show that the traces that start at
  $\mathtt{rand3}$ and end at $[\retkw_{\intT} \_] \cdot n$ have
  respective weights $1/4$, $1/4 \cdot 1/4$, \ldots,
  $(1/4)^n \cdot 1/4$, \ldots, and that the sum of those weights is
  $1/3$.
\end{example}

The following is immediate.
\begin{lemma}
  \label{lemma:Prob:compute}
  The following hold:
  \begin{enumerate}
  \item For every rule $s \step \alpha t$, $t$ and $s$ have the same
    type.
  \item For every rule of the form $s \step 1 t$ of type $D\intT$, for
    every $n \in \Z$,
    $\Prob [t \vp n] = \Prob [s \vp n]$.
  \item
    $\Prob [C \cdot M \oplus N \vp n] = \frac 1 2 \Prob [C \cdot M \vp
    n] + \frac 1 2 \Prob [C \cdot N \vp n]$.  \qed
  \end{enumerate}
\end{lemma}

\section{A refresher on domain theory}
\label{sec:refr-doma-theory}

We will require some elementary domain theory, for which we refer the
reader to \cite{GHKLMS:contlatt,AJ:domains,JGL-topology}.  A
\emph{poset} $X$ is a set with a partial ordering, which we will
always write as $\leq$.  A \emph{directed family} $D \subseteq X$ is a
non-empty family such that every pair of points of $D$ has an upper
bound in $D$.  A \emph{dcpo} is a poset in which every directed family
$D$ has a supremum $\dsup D$.  If $D = {(x_i)}_{i \in I}$, we also
write $\dsup_{i \in I} x_i$ for $\dsup D$.

The \emph{product} $X \times Y$ of two dcpos is the set of pairs $(x,
y)$, $x \in X$, $y \in Y$, ordered by $(x, y) \leq (x', y')$ if and
only if $x \leq x'$ and $y \leq y'$.

For any two dcpos $X$ and $Y$, a map $f \colon X \to Y$ is
\emph{Scott-continuous} if and only if it is monotonic ($x \leq x'$
implies $f (x) \leq f (x')$) and preserves directed suprema (for every
directed family ${(x_i)}_{i \in I}$ in $X$,
$\dsup_{i \in I} f (x_i) = f (\dsup_{i \in I} x_i)$).  There is a
category $\Dcpo$ of dcpos and Scott-continuous maps.

We order maps from $X$ to $Y$ by $f \leq g$ if and only if
$f (x) \leq g (x)$ for every $x \in X$.  The poset $[X \to Y]$ of all
Scott-continuous maps from $X$ to $Y$ is then again a dcpo, and
directed suprema are computed pointwise:
$(\dsup_{i \in I} f_i) (x) = \dsup_{i \in I} (f_i (x))$.  $\Dcpo$ is a
Cartesian-closed category---a model of simply-typed
$\lambda$-calculus---and that can be said more concretely as follows:
\begin{itemize}
\item for all dcpos $X$, $Y$, there is a Scott-continuous map $\App
  \colon  [X \to Y] \times X \to Y$ defined by $\App (f, x) \eqdef f
  (x)$;
\item for all dcpos $X$, $Y$, $Z$, for every Scott-continuous map $f
  \colon Z \times X \to Y$, the map $\Lambda_X (f) \colon Z \to [X \to
  Y]$ defined by $\Lambda_X (f) (z) (x) \eqdef f (z, x)$ is
  Scott-continuous;
\item those satisfy certain equations which we will not require.
\end{itemize}

If the dcpo $X$ is \emph{pointed}, namely if it has a least element
$\bot$, then every Scott-continuous map $f \colon X \to X$ has a least
fixed point $\lfp_X (f) \eqdef \dsup_{n \in \nat} f^n (\bot)$.  This
is used to interpret recursion.  Additionally, the map
$\lfp_X \colon [X \to X] \to X$ is itself Scott-continuous.

The set $\creal \eqdef \Rplus \cup \{\infty\}$ of extended
non-negative real numbers is a dcpo under the usual ordering.  We
write $\Lform X$ for $[X \to \creal]$.  Its elements are called the
\emph{lower semicontinuous} functions in analysis.

A \emph{Scott-open} subset $U$ of a dcpo $X$ is an upwards-closed
subset ($x \in U$ and $x \leq y$ imply $y \in U$) that is inaccessible
from below (every directed family $D$ such that $\dsup D \in U$
intersects $U$).  The lattice of Scott-open subsets is written
$\Open X$, and forms a topology, the \emph{Scott topology} on $X$.
Note that $\Open X$ is itself a dcpo under inclusion, and directed
suprema are computed as unions.

The \emph{Scott-closed} sets are the complements of Scott-open sets,
i.e., the down\-wards-closed subsets $C$ such that for every directed
family $D \subseteq C$, $\dsup D \in C$.

In order to give a denotational semantics to probabilistic choice, we
will follow Jones \cite{jones89,Jones:proba}.  A \emph{continuous
  valuation} on $X$ is a map $\nu \colon \Open X \to \creal$ that is
\emph{strict} ($\nu (\emptyset)=0$), \emph{monotone} ($U \subseteq V$
implies $\nu (U) \leq \nu (V)$), \emph{modular}
($\nu (U) + \nu (V) = \nu (U \cup V) + \nu (U \cap V)$), and
Scott-continuous
($\nu (\dcup_{i \in I} U_i) = \dsup_{i \in I} \nu (U_i)$).  A
\emph{subprobability valuation} additionally satisfies
$\nu (X) \leq 1$.  Continuous valuations and measures are very close
concepts: see \cite{KL:measureext} for details.

Among subprobability valuations, one finds the \emph{Dirac valuation}
$\delta_x$, for each $x \in X$, defined by $\delta_x (U) \eqdef 1$ if
$x \in U$, $0$ otherwise.  One can integrate any Scott-continuous map
$f \colon X \to \creal$, and the integral $\int_{x \in X} f (x) d\nu$
is Scott-continuous and linear (i.e., commutes with sums and scalar
products by elements of $\Rplus$) both in $f$ and in $\nu$.

We write $\Val_{\leq 1} X$ for the poset of subprobability valuations
on $X$.  This is a dcpo under the pointwise ordering ($\mu \leq \nu$
if and only if $\mu (U) \leq \nu (U)$ for every $U \in \Open X$), and
directed suprema are computed pointwise
($(\dsup_{i \in I} \nu_i) (U) = \dsup_{i \in I} (\nu_i (U))$).
Additionally, $\Val_{\leq 1}$ defines a \emph{monad} on $\Dcpo$.
Concretely:
\begin{itemize}
\item there is a \emph{unit} $\eta \colon X \to \Val_{\leq 1} X$,
  which is the continuous map $x \mapsto \delta_x$;
\item every Scott-continuous map $f \colon X \to \Val_{\leq 1} Y$ has
  an \emph{extension} $f^\dagger \colon \Val_{\leq 1} X \to \Val_{\leq
    1} Y$, defined by $f^\dagger (\nu) (V) \eqdef \int_{x \in X} f (x)
  (V) d\nu$;
\item those satisfy a certain number of equations, of which we will
  need the following:
  \begin{align}
    \label{eq:dagger:eta}
    f^\dagger (\eta (x)) & = f (x) \\
    \label{eq:dagger:dbl}
    \int_{y \in Y} h (y) d f^\dagger (\nu) & = \int_{x \in X} \left(
                                             \int_{y \in Y} h (y) df
                                             (x)
                                             \right) d\nu,
  \end{align}
  for all Scott-continuous maps $f \colon X \to Y$, $h \colon Y \to
  \creal$, and every $\nu \in \Val_{\leq 1} X$.
\end{itemize}
Note that the map $f \mapsto f^\dagger$ is itself Scott-continuous.

\section{Denotational semantics}
\label{sec:denot-semant}

The types $\tau$ are interpreted as dcpos $\Eval \tau$, as follows:
$\Eval {\intT} \eqdef \Z$, with equality as ordering;
$\Eval {D \tau} \eqdef \Val_{\leq 1} \Eval \tau$; and
$\Eval {\sigma \to \tau} \eqdef [\Eval \sigma \to \Eval \tau]$.  Note
that $\Eval \tau$ is pointed for every computation type $\tau$, so
$\lfp_{\Eval \tau}$ makes sense in those cases.

An \emph{environment} is a map $\rho$ sending each variable $x_\tau$
to an element $\rho (x_\tau)$ of $\Eval \tau$.  The dcpo $Env$ of
environments is the product
$\prod_{x_\tau \text{ variable}} \Eval \tau$, with the usual
componentwise ordering.  When $V \in \Eval \sigma$, we write
$\rho [x_\sigma := V]$ for the environment that maps $x_\sigma$ to
$V$, and all other variables $y$ to $\rho (y)$.

\begin{figure}
  \centering
  \[
  \begin{array}{c@{\quad}c}
    \Eval {x_\tau} \rho \eqdef \rho (x_\tau) \quad
    \Eval n \rho \eqdef n \ (n \in \Z) &
    \Eval {\suc M} \rho \eqdef \Eval M \rho + 1 \quad
                                                  \Eval {\p M} \rho \eqdef \Eval M \rho - 1 \\
    \multicolumn{2}{c}{
    \Eval {\ifz M N P} \rho \eqdef \left\{
                              \begin{array}{ll}
                                \Eval N \rho & \text{if } \Eval M \rho=0
                                \\
                                \Eval P \rho & \text{otherwise}
                              \end{array}
                                               \right.}
    \\
    \Eval {MN} \rho \eqdef \App (\Eval M \rho, \Eval N \rho)
                                       &
    \Eval {\lambda x_\sigma . M} \rho \eqdef
                                        (V \in \Eval \sigma
                                        \mapsto \Eval M \rho
                                        [x_\sigma:=V]) \\
    \Eval {\reckw_\tau M} \rho \eqdef \lfp_{\Eval \tau} (\Eval M \rho)
    &
    \Eval {M \oplus N} \rho \eqdef \frac 1 2 (\Eval M \rho + \Eval N
                              \rho) \\
    \Eval {\retkw_\sigma M} \rho \eqdef \eta (\Eval M \rho) =
                                   \delta_{\Eval M \rho} &
    \Eval {\bindkw_{\sigma,\tau} M N} \rho \eqdef (\Eval N
                                             \rho)^\dagger (\Eval M \rho).
  \end{array}
  \]
  \caption{Denotational semantics}
  \label{fig:denot}
\end{figure}

Let us write $V \in X \mapsto f (V)$ for the function that maps every
$V \in X$ to the value $f (V)$.  We can now define the value $\Eval M$
of terms $M \colon \tau$, as Scott-continuous maps
$\rho \in Env \mapsto \Eval M \rho$, by induction on $M$, see
Figure~\ref{fig:denot}.

The operational semantics and the denotational
semantics match, namely:
\begin{theorem}[Adequacy]
  \label{thm:R:adeq}
  For every ground term $M \colon D\intT$, for every $n \in \Z$,
  $\Eval M (\{n\}) = \Prob [M \vp n]$.
\end{theorem}
\ifspringer%
The proof is relatively standard, using appropriate logical relations,
and is given in the full version of
this paper \cite[Appendices]{JGL:8/27}.
\else%
The proof is relatively standard, and given in the appendices.
Appendix~\ref{sec:soundness} establishes soundness, namely
$\Eval M (\{n\}) \geq \Prob [M \vp n]$, and
Appendix~\ref{sec:adequacy} shows the converse inequality, using
appropriate logical relations.
\fi

\begin{example}
  \label{ex:randnat:denot}
  We retrieve the result of Example~\ref{ex:randnat:op} using adequacy
  as follows.
  $\Eval {\lambda r_{\intT \to D\intT} . \lambda m_{\intT} . r (\suc
    m) \oplus \retkw_{\intT} m}$ is the function $F$ that maps every
  $\varphi \in \Eval {\intT \to D\intT}$ (the value of $r$) and every
  $m \in \Eval {\intT} = \Z$ to $1/2 \varphi (m+1) + 1/2 \delta_m$.
  Let $\varphi_k \eqdef F^k (\bot)$, for every $k \in \nat$.  Then
  $\varphi_0 = \bot$ maps every $m \in \nat$ to the zero valuation
  $0$, $\varphi_1 (m) = 1/2 \delta_m$ for every $m \in \nat$,
  $\varphi_2 (m) = 1/4 \delta_{m+1} + 1/2 \delta_m$ for every
  $m \in \nat$, etc.  By induction on $k$,
  $\varphi_k (m) = \sum_{i=0}^{k-1} 1/2^{i+1} \delta_{m+i}$.  Taking
  suprema over $k$, we obtain that
  $\lfp_{\Eval {\intT \to D\intT}} (F)$ maps every $m \in \nat$ to
  $\sum_{i=0}^\infty 1/2^{i+1} \delta_{m+i}$.  Then
  $\Eval {\mathtt{rand\_nat}} = \lfp_{\Eval {\intT \to D\intT}} (F)
  (0) = \sum_{n \in \nat} \frac 1 {2^{n+1}} \delta_n$.
\end{example}

\begin{example}
  \label{ex:rand3:denot}
  We retrieve the result of Example~\ref{ex:rej:sample:3}, using
  adequacy, as follows.  The semantics of
  $\lambda p_{D\intT} . ((\retkw_\intT 0 \oplus \retkw_\intT 1) \oplus
  (\retkw_\intT 2 \oplus p_{D\intT}))$ is the function $f$ that maps
  every $\nu \in \Eval {D\intT}$ to
  $\frac 1 4 \delta_0 + \frac 1 4 \delta_1 + \frac 1 4 \delta_2 +
  \frac 1 4 \nu$.  For every $n \in \nat$,
  $f^n (0) = a_n \delta_0 + a_n \delta_1 + a_n \delta_2$ where
  $a_n = 1/4 + (1/4)^2 + \cdots + (1/4)^n = 1/4 (1 - (1/4)^n) /
  (1-1/4)$.  Since $\Eval {\intT}$ has equality as ordering, the
  ordering on $\Eval {D\intT}$ is given by comparing the coefficients
  of each $\delta_N$, $N \in \Eval {\intT}$.  In particular, the least
  fixed point of $f$ is obtained as
  $a \delta_0 + a \delta_1 + a \delta_2$, where
  $a \eqdef \dsup_{n \in \nat} a_n = 1/3$.
\end{example}

\begin{example}
  \label{ex:rej:sample}
  Here is a lengthier example, which we will leave to the reader.
  While lengthy, working denotationally is doable.  Proving the same
  argument operational would be next to impossible, even in the
  special case $\tau = \intT$.
  
  We define a more general form of rejection sampling, as follows.
  Let $\tau$ be any type.  We consider the PCF$_\Nature$ term:
  \begin{align*}
    \mathtt{sample}
    & \eqdef \lambda p_{D\tau} . \lambda sel_{\tau \to D\intT} . \\
    & \qquad\reckw_{D\tau} (\lambda r_{D\tau} .
      \dokw {x_\tau \leftarrow p_{D\tau}}; \\
    & \qquad\qquad\qquad\quad \dokw {b_\intT \leftarrow sel_{\tau \to D\intT}
      x_\tau}; \\
    & \qquad\qquad\qquad\quad \ifz {b_\boolT} {\retkw_\tau x_\tau} {r_{D\tau}}).
  \end{align*}
  The idea is that we draw $x$ according to distribution $p$, then we
  call $sel$ as a predicate on $x$.  If the result, $b$, is true
  (zero) then we return $x$, otherwise we start all over.  Note that
  $sel$ can itself return a \emph{random} $b$.
  
  For every $g \in \Lform {\Eval \tau}$, and every
  $\nu \Eval {D\tau}$, we let $g \cdot \nu$ (sometimes written
  $g \; d\nu$) be the continuous valuation defined from $\nu$ by using
  $g$ as a density, namely
  $(g \cdot \nu) (U) \eqdef \int_{x \in \Eval \tau} \chi_U (x) g (x)
  d\nu$ for every open subset $U$ of $\Eval \tau$, where $\chi_U$ is
  the characteristic map of $U$.  One can check that
  $g \cdot \nu = (x \mapsto g (x) \delta_x)^\dagger (\nu)$, using the
  equality $\chi_U (x) = \delta_x (U)$, and, using
  (\ref{eq:dagger:dbl}), that for every $h \in \Lform {\Eval \tau}$,
  $\int_{x \in X} h (x) d (g \cdot \nu) = \int_{x \in X} h (x) g (x) d
  \nu$.

  For every $s \in \Eval {\tau \to D\intT}$, for every
  $x \in \Eval \tau$, let $s_0 (x) \eqdef s (x) (\{0\})$,
  $s_1 (x) \eqdef s (x) (\Z \diff \{0\})$.  We let the reader check
  that, for every environment $\rho$, $\Eval {\mathtt{sample}} \rho$
  maps every subprobability valuation $\nu$ on $\Eval \tau$ and every
  $s \in \Eval {\tau \to D\intT}$ to the subprobability valuation
  $\frac 1 {1 - (s_1 \cdot \nu) (\Eval \tau)} (s_0 \cdot \nu)$ if
  $(s_1 \cdot \nu) (\Eval \tau) \neq 1$, to the zero valuation
  otherwise.

  In particular, if $s$ is a predicate, implemented as a function that
  maps every $x \in U \subseteq \Eval \tau$ to $\delta_0$ and every
  $x \in V \subseteq \Eval \tau$ (for some disjoint open sets $U$ and
  $V$) to $\delta_1$, so that $s_0 = \chi_U$ and $s_1 = \chi_V$,
  then $\Eval {\mathtt{sample}} \rho (\nu) (s)$ is the subprobability
  valuation $\frac 1 {1-\nu (V)} \nu_{|U}$ if $\nu (V) \neq 1$, the
  zero valuation otherwise.  ($\nu_{|U}$ denotes the restriction of
  $\nu$ to $U$, defined by $\nu_{|U} (V) \eqdef \nu (U \cap V)$.)

  In the special case where $V$ is the complement of $U$, it follows
  that $\mathtt{sample}$ implements \emph{conditional probabilities}:
  $\Eval {\mathtt{sample}} \rho (\nu) (s) (W)$ is the probability that
  a $\nu$-random element lies in $W$, conditioned on the fact that it
  is in $U$.
\end{example}

\section{The parallel or tester}
\label{sec:parallel-or-tester}

In PCF$_\Nature$, computation happens at type $D\intT$, not $\intT$,
hence let us call \emph{parallel or} function any
$f \in \Eval {D\intT \to D\intT \to \allowbreak D\intT}$ such that
$f (\delta_1) (\delta_1) = \delta_1$ and
$f (\delta_0) (\nu) = f (\nu) (\delta_0) = \delta_0$ for every
$\nu \in \Eval {D\intT}$.  Realizing that every element of
$\Eval {D\intT}$ is of the form $a \delta_0 + b \delta_1$, with
$a, b \in \Rplus$ such that $a+b \leq 1$, the function $por$ defined
by
$por (a \delta_0 + b \delta_1) (a' \delta_0 + b' \delta_1) \eqdef
(a+a'-aa') \delta_0 + bb' \delta_1$ is such a parallel or function.

Note how parallel ors differ from the usual \emph{left-to-right
  sequential or} used in most programming languages:
\begin{align*}
  \mathtt{lror} & \eqdef \lambda p_{D\intT} . \lambda q_{D\intT}
                  . \\
  & \dokw {x_\intT \leftarrow p_{D\intT}};
    \ifz {x_\intT} {\retkw_\intT 0} {q_{D\intT}}
\end{align*}
whose semantics is given by
$\Eval {\mathtt{lror}} (a \delta_0 + b \delta_1) (a' \delta_0 + b'
\delta_1) = (a+ba') \delta_0 + bb' \delta_1$---so
$\Eval {\mathtt{lror}}$ maps $\delta_1, \delta_1$ to $\delta_1$, and
$\delta_0,\nu$ to $\delta_0$, but maps
$a \delta_0 + b \delta_1, \delta_0$ to $(a+b) \delta_0$, not
$\delta_0$.  Symmetrically, there is a \emph{right-to-left sequential
  or}:
\begin{align*}
  \mathtt{rlor} & \eqdef \lambda p_{D\intT} . \lambda q_{D\intT} .\\
  & \dokw {x_\intT \leftarrow q_{D\intT}};
    \ifc {x_\intT} {\retkw_\intT 0} {p_{D\intT}}.
\end{align*}

We define a \emph{parallel or tester} as follows:
\begin{align*}
  \mathtt{portest}
  & \eqdef \lambda f_{D\intT \to D\intT \to D\intT} . \\
  & \dokw {x_\intT \leftarrow f (\retkw_{\intT} 1)
    (\retkw_{\intT} 1)}; \\
  & \ifkw {x_\intT=0}\;\mathtt{then}\; {\Omega} \\
  & \elsekw (\dokw {y_\intT \leftarrow f (\retkw_{\intT} 0)
    (\Omega)}; \\
  & \qquad\ \ \ifkw {y_\intT=0}\;\mathtt{then}\;(\dokw {z_\intT
    \leftarrow f (\Omega) (\retkw_{\intT} 0)}; \\
  & \qquad\qquad\qquad\qquad\qquad\ \ifz {z_\intT} {\retkw_{\unitT} 0}
    {\Omega}) \\
  & \qquad\ \ \elsekw \Omega),
\end{align*}
where $\Omega \eqdef \recbe (\lambda a_{D\intT} . a_{D\intT})$.  One
can check that $\Eval {\mathtt{portest}} (por) = \delta_{0}$, and that
would hold for any other parallel or function instead of $por$.  If
things worked in PCF$_\Nature$ as in PCF, we would be able to show
that $\mathtt{portest}$ is contextually equivalent to the constant map
that loops on every input $f_{D\intT \to D\intT \to D\intT}$.

However, that is not the case.  As we will now see, there is a
PCF$_\Nature$ term, the \emph{poor man's parallel or}
$\mathtt{pmpor}$, such that $\mathtt{portest}\; \mathtt{pmpor}$
terminates with non-zero probability.  That term takes its two
arguments of type $D\intT$, then decides to do one of the following
three actions with equal probability $1/3$: (1) call $\mathtt{lror}$
on the two arguments; (2) call $\mathtt{rlor}$ on the two arguments;
or (3) return true ($0$), regardless of its arguments.

In order to define $\mathtt{pmpor}$, we need to draw an element out of
three with equal probability.  We do that by rejection sampling,
imitating $\mathtt{rand3}$ (Examples~\ref{ex:rej:sample:3},
\ref{ex:rand3:op} and \ref{ex:rand3:denot}): we draw one element among
four with equal probability, and we repeat until it falls in a
specified subset of three.  Hence we define:
\begin{align*}
  \mathtt{pmpor}
  & \eqdef \lambda p_{D\intT} . \lambda q_{D\intT} .
    \reckw_{D\intT} (\lambda r . \\
  & ((\mathtt{lror} \; p \; q) \oplus (\mathtt{rlor} \; p \; q)) \oplus
    (\retkw_{\intT} 0 \oplus r))
\end{align*}
One can show that $\Eval {\mathtt{pmpor}}$ maps every pair of
subprobability distributions $\mu$, $\nu$ on $\Eval {\intT}$ to
$\frac 1 3 \Eval {\mathtt{rlor}} (\mu) (\nu) + \frac 1 3 \Eval
{\mathtt{lror}} (\mu) (\nu) + \frac 1 3 \delta_0$.  Intuitively,
$\mathtt{portest}\; \mathtt{pmpor}$ will terminate with probability
$(2/3)^3 = 8/27 \approx 0.296296\ldots$: with $f = \mathtt{pmpor}$,
the first test $f (\delta_1) (\delta_1) = \delta_1$
will succeed whether $f$ acts as $\mathtt{lror}$ or as $\mathtt{rlor}$
(but not as the constant map returning $\delta_0$), which happens
with probability $2/3$; the second test
$f (\delta_0) (0) = \delta_0$ will succeed whether $f$ acts as
$\mathtt{lror}$ or as the constant map returning $\delta_0$ (but
not as $\mathtt{rlor}$), again with probability $2/3$; and the final
test $f (0) (\delta_0) = \delta_0$ will symmetrically succeed
with probability $2/3$.

We now show that the probability $8/27$ is optimal.  To this end, we
need to use a logical relation $(\tr_\tau)_{\tau \text{ type}}$,
namely a family of relations $\tr_\tau$, one for each type $\tau$, and
related by certain constraints to be described below.  Each $\tr_\tau$
will be an $I$-ary relation on values in $\Eval \tau$, for some
non-empty set $I$, namely $\tr_\tau \subseteq (\Eval \tau)^I$.  In
practice, we will take $I \eqdef \{1, 2, 3\}$, but the proofs are
easier if we keep $I$ arbitrary for now.

Our construction will be parameterized by an $I$-ary relation
$\tr \subseteq \creal^I$.  We will also define an auxiliary family of
relations $\tr_\tau^\perp$, as certain subsets of
$(\Lform {\Eval \tau})^I$.  We require $\tr$ to contains the all zero
tuple $\vec 0 \eqdef {(0)}_{i \in I}$, to be closed under directed
suprema, and to be convex.  (By \emph{convex}, we mean that for all
$\vec x, \vec y \in \tr$ and $a \in [0, 1]$, $a \vec x + (1-a) \vec y$
is in $\tr$ as well.)

We define:
\begin{itemize}
\item ${(n_i)}_{i \in I} \in \tr_\intT$ if and only if all $n_i$ are
  equal;
\item ${(f_i)}_{i \in I} \in \tr_{\sigma \to \tau}$ if and only if for all
  ${(V_i)}_{i \in I} \in \tr_\sigma$, ${(f_i (V_i))}_{i \in I} \in \tr_\tau$;
\item ${(\nu_i)}_{i \in I} \in \tr_{D\tau}$ if and only if for all
  ${(h_i)}_{i \in I} \in \tr_\tau^\perp$,
  ${(\int_{V \in \Eval \tau} h_i (V) d\nu_i)}_{i \in I} \in \tr$;
\item ${(h_i)}_{i \in I} \in \tr_\tau^\perp$ if and only if for all
  ${(V_i)}_{i \in I} \in \tr_\tau$, ${(h_i (V_i))}_{i \in I} \in \tr$.
\end{itemize}
We also define $\tr_* \subseteq Env^I$ by
${(\rho_i)}_{i \in I} \in \tr_*$ if and only if for every variable
$x_\sigma$, ${(\rho_i (x_\sigma))}_{i \in I} \in \tr_\sigma$.  We
prove the following \emph{basic lemma of logical relations}:
\begin{proposition}
  \label{prop:I:basic}
  For all ${(\rho_i)}_{i \in I} \in \tr_*$, for every $M \colon \tau$,
  ${(\Eval M \rho_i)}_{i \in I}$ is in $\tr_\tau$.
\end{proposition}
\begin{proof}
  Step~1.  We claim that for every type $\tau$, $\tr_\tau$ is closed
  under directed suprema taken in ${(\Eval \tau)}^I$, and contains the
  least element ${(\bot_\tau)}_{i \in I}$ if $\tau$ is a computation
  type.  This is by induction on $\tau$.  The claim is trivial for
  $\intT$, since $\Eval {\intT}^I$ is ordered by equality.  For every
  directed family ${(\vec f_j)}_{j \in J}$ in $\tr_{\sigma \to \tau}$,
  with $\vec f_j \eqdef {(f_{ji})}_{i \in I}$, we form its supremum
  $\vec f \eqdef {(f_i)}_{i \in I}$ pointwise, namely
  $f_i \eqdef \dsup_{j \in J} f_{ji}$.  For every
  ${(V_i)}_{i \in I} \in \tr_\sigma$, ${(f_{ji} (V_i))}_{i \in I}$ is
  in $\tr_\tau$ for every $j \in J$, so by induction hypothesis
  ${(f_i (V_i))}_{i \in I}$ is also in $\tr_\tau$.  It follows that
  ${(f_i)}_{i \in I}$ is in $\tr_{\sigma \to \tau}$.  For every
  directed family ${(\vec \nu_j)}_{j \in J}$ in $\tr_{D\tau}$, with
  $\vec \nu_j \eqdef {(\nu_{ji})}_{i \in I}$, we form its supremum
  $\vec \nu \eqdef {(\nu_i)}_{i \in I}$ pointwise, that is
  $\nu_i \eqdef \dsup_{j \in J} \nu_{ji}$.  For all
  ${(h_i)}_{i \in I} \in \tr_\tau^\perp$,
  ${(\int_{V \in \Eval \tau} h_i (V) d\nu_{ji})}_{i \in I} \in \tr$
  for every $j \in J$, by induction hypothesis.  We take suprema over
  $j \in J$.  Since $\tr$ is closed under directed suprema, and
  integration is Scott-continuous in the valuation,
  ${(\int_{V \in \Eval \tau} h_i (V) d\nu_i)}_{i \in I}$ is in $\tr$.
  Since ${(h_i)}_{i \in I}$ is arbitrary, ${(\nu_i)}_{i \in I} \in \tr_{D\tau}$.

  We also show that ${(\bot_\tau)}_{i \in I} \in \tr_\tau$ for every
  computation type $\tau$.  For function types, this is immediate.
  For types of the form $D\tau$, we must check that $\vec 0$ is in
  $\tr_{D\tau}$.  For all ${(h_i)}_{i \in I} \in \tr_\tau^\perp$, we
  indeed have
  ${(\int_{V \in \Eval \tau} h_i (V) d0)}_{i \in I} \in \tr$, since
  $\vec 0 \in \tr$.

  Step 2.  We claim that for all
  ${(\nu_i)}_{i \in I} \in \tr_{D\sigma}$, for all
  ${(f_i)}_{i \in I} \in \tr_{\sigma \to D\tau}$,
  ${(f_i^\dagger (\nu_i))}_{i \in I} \in \tr_{D\tau}$.  
  We wish to use the definition of $\tr_{D\tau}$, so we consider an
  arbitrary tuple ${(h_i)}_{i \in I} \in \tr_\tau^\perp$, and we aim
  to prove that
  ${(\int_{V \in \Eval \tau} h_i (V) df_i^\dagger (\nu_i))}_{i \in I}$
  is in $\tr$.  For that, we use equation~(\ref{eq:dagger:dbl})%
  \ifspringer%
  : %
  \else%
  , to the effect that %
  \fi%
  $\int_{V \in \Eval \tau} h_i (V) df_i^\dagger (\nu_i) = \int_{x \in
    \Eval \sigma} \left(\int_{V \in \Eval \tau} h_i (V) df_i
    (x)\right) d\nu_i$, for every $i \in I$.

  Let us define
  $h'_i (x) \eqdef \int_{V \in \Eval \tau} h_i (V) df_i (x)$.  We
  claim that ${(h'_i)}_{i \in I} \in \tr_\sigma^\perp$.  Let
  ${(x_i)}_{i \in I} \in \tr_\sigma$.  Then
  ${(f_i (x_i))}_{i \in I} \in \tr_{D\tau}$, and since
  ${(h_i)}_{i \in I} \in \tr_\tau^\perp$, ${(h'_i (x_i))}_{i \in I}$
  is in $\tr$, by definition of $\tr_{D\tau}$.  Since
  ${(x_i)}_{i \in I}$ is arbitrary,
  ${(h'_i)}_{i \in I} \in \tr_\sigma^\perp$.

  Since ${(h'_i)}_{i \in I} \in \tr_\sigma^\perp$ and
  ${(\nu_i)}_{i \in I} \in \tr_{D\sigma}$, by definition of
  $\tr_{D\sigma}$ we obtain that
  ${(\int_{x_i \in \Eval {D\sigma}} h'_i (x_i) d\nu_i)}_{i \in I}$ is
  in $\tr$, and this is exactly what we wanted to prove.

  We now prove the claim by induction on $M$.  If $M$ is a variable,
  this is by assumption.  If $M=0$, this is trivial.  If $M$ is of the
  form $\suc N$, then all the values $\Eval N \rho_i$ are equal, hence
  also all the values $\Eval M \rho_i = \Eval N \rho_i + 1$.
  Similarly for terms of the form $\p N$.  The case of applications is
  by definition of $\tr_{\sigma \to \tau}$.  In the case of
  abstractions $\lambda x_\sigma . M$ with $M \colon \tau$, we must
  show that, letting $f_i$ be the map
  $V \in \Eval \sigma \mapsto \Eval M (\rho_i [x_\sigma \mapsto V])$
  ($i \in I$), for all ${(V_i)}_{i \in I} \in \tr_\sigma$,
  ${(f_i (V_i))}_{i \in I} \in \tr_\tau$.  This boils down to checking
  that
  ${(\Eval M (\rho_i [x_\sigma \mapsto V_i]))}_{i \in I} \in \tr_\tau$
  for all ${(V_i)}_{i \in I} \in \tr_\sigma$, which follows
  immediately from the induction hypothesis and the easily checked
  fact that ${(\rho_i [x_\sigma \mapsto V_i])}_{i \in I}$ is in
  $\tr_*$.

  The case of terms of the form $\reckw_\tau M$, where $\tau$ is a
  computation type, is more interesting.  Let $f_i$ be the map
  $\Eval M \rho_i \colon \Eval \tau \to \Eval \tau$.  By induction
  hypothesis ${(f_i)}_{i \in I}$ is in $\tr_{\tau \to \tau}$, so for
  all ${(a_i)}_{i \in I} \in \tr_\tau$, ${(f_i (a_i))}_{i \in I}$ is
  in $\tr_\tau$.  Iterating this, we have
  ${(f_i^n (a_i))}_{i \in I} \in \tr_\tau$ for every $n \in \nat$.  By
  Step~1, ${(\bot_\tau)}_{i \in I}$ is in $\tr_\tau$.  Hence
  ${(f_i^n (\bot_\tau))}_{i \in I} \in \tr_\tau$ for every
  $n \in \nat$.  Since $\tr_\tau$ is closed under directed suprema by
  Step~1,
  ${(\lfp_{\Eval \tau} f_i)}_{i \in I} = {(\Eval {\reckw_\tau M}
    \rho_i)}_{i \in I}$ is in $\tr_\tau$.

  For terms of the form $M \eqdef \ifz N P Q$ of type $\tau$, by
  induction hypothesis ${(\Eval N \rho_i)}_{i \in I} \in \tr_{\intT}$,
  so all values $\Eval N \rho_i$ are the same integer, say $n$.  (And
  this term exists because $I$ is non-empty.)  If $n=0$, then for
  every $i \in I$, $\Eval M \rho_i$ is then equal to $\Eval P \rho_i$,
  so ${(\Eval M \rho_i)}_{i \in I} = {(\Eval P \rho_i)}_{i \in I}$ is
  in $\tr_\tau$.  We reason similarly if $n \neq 0$.

  For terms of the form $M \oplus N$, of type $D\tau$, we consider an
  arbitrary tuple ${(h_i)}_{i \in I} \in \tr_\tau^\perp$.  By
  induction hypothesis ${(\Eval M \rho_i)}_{i \in I}$ and
  ${(\Eval N \rho_i)}_{i \in I}$ are in $\tr_{D\tau}$, so
  ${(\int_{V \in \Eval \tau} h_i (V) d\Eval M\rho_i)}_{i \in I}$ and
  ${(\int_{V \in \Eval \tau} h_i (V) d\Eval N\rho_i)}_{i \in I}$ are
  in $\tr$.  Since $\tr$ is convex, and integration is linear in the
  valuation,
  ${(\int_{V \in \Eval \tau} h_i (V) d\Eval {M \oplus N}\rho_i)}_{i
    \in I}$ is also in $\tr$.  Since ${(h_i)}_{i \in I}$ is arbitrary,
  ${(\Eval {M \oplus N} \rho_i)}_{i \in I}$ is in $\tr_{D\tau}$.

  For terms of the form $\retkw_\sigma M$, we again consider an
  arbitrary tuple ${(h_i)}_{i \in I}$ in $\tr_\sigma^\perp$.  By
  induction hypothesis, ${(\Eval M \rho_i)}_{i \in I}$ is in
  $\tr_\sigma$, so by definition of $\tr_\sigma^\perp$,
  ${(h_i (\Eval M \rho_i))}_{i \in I}$ is in $\tr$.  Equivalently,
  ${(\int_{V \in \Eval \sigma} h_i (V) d\delta_{\Eval M\rho_i})}_{i
    \in I}$ is in $\tr$, and that means that
  ${(\Eval {\retkw_\sigma M} \rho_i)}_{i \in I}$ is in
  $\tr_{D\sigma}$.

  Finally, for terms $\bindkw_{\sigma, \tau} M N$, we have
  ${(\Eval M \rho_i)}_{i \in I} \in \tr_{D\sigma}$ and
  ${(\Eval N \rho_i)}_{i \in I} \in \tr_{\sigma \to D\sigma}$ by
  induction hypothesis, so
  ${(\Eval {\bindkw_{\sigma, \tau} M N} \rho_i)}_{i \in I} \in \tr_{D
    \tau}$by Step~2.  \qed
\end{proof}

\begin{proposition}
  \label{prop:portest:8/27}
  For every ground PCF$_\Nature$ term
  $P \colon D\intT \to D\intT \to D\intT$,
  $\Eval {\mathtt{portest}\;P} \leq 8/27 \cdot \delta_{0}$.
\end{proposition}
\begin{proof}
  We specialize the construction of the logical relation
  ${(\tr_\tau)}_{\tau \text{ type}}$ 

  \noindent
  \begin{minipage}{0.62\linewidth}
    to $I \eqdef \{1, 2, 3\}$ and to $\tr$, defined as the downward
    closure in $\Rplus^3$ of the convex hull
    $\{a \cdot (1, 0, 1) + b \cdot (1, 1, 0) + c \cdot (0, 1, 1) \mid
    a, b, c \in \Rplus, a+b+c \leq 1\}$ of the three points
    $\vec \alpha_1 \eqdef (1, 0, 1)$,
    $\vec \alpha_2 \eqdef (1, 1, 0)$, and
    $\vec \alpha_3 \eqdef (0, 1, 1)$.  The relation $\tr$ has an
    alternate description as the set of those points $(a, b, c)$ of
    $\Rplus^3$ such that $a, b, c \leq 1$ and $a+b+c \leq 2$.  This is
    depicted on the right.
  \end{minipage}
  \begin{minipage}{0.37\linewidth}
    \begin{flushright}
      \includegraphics[scale=0.3]{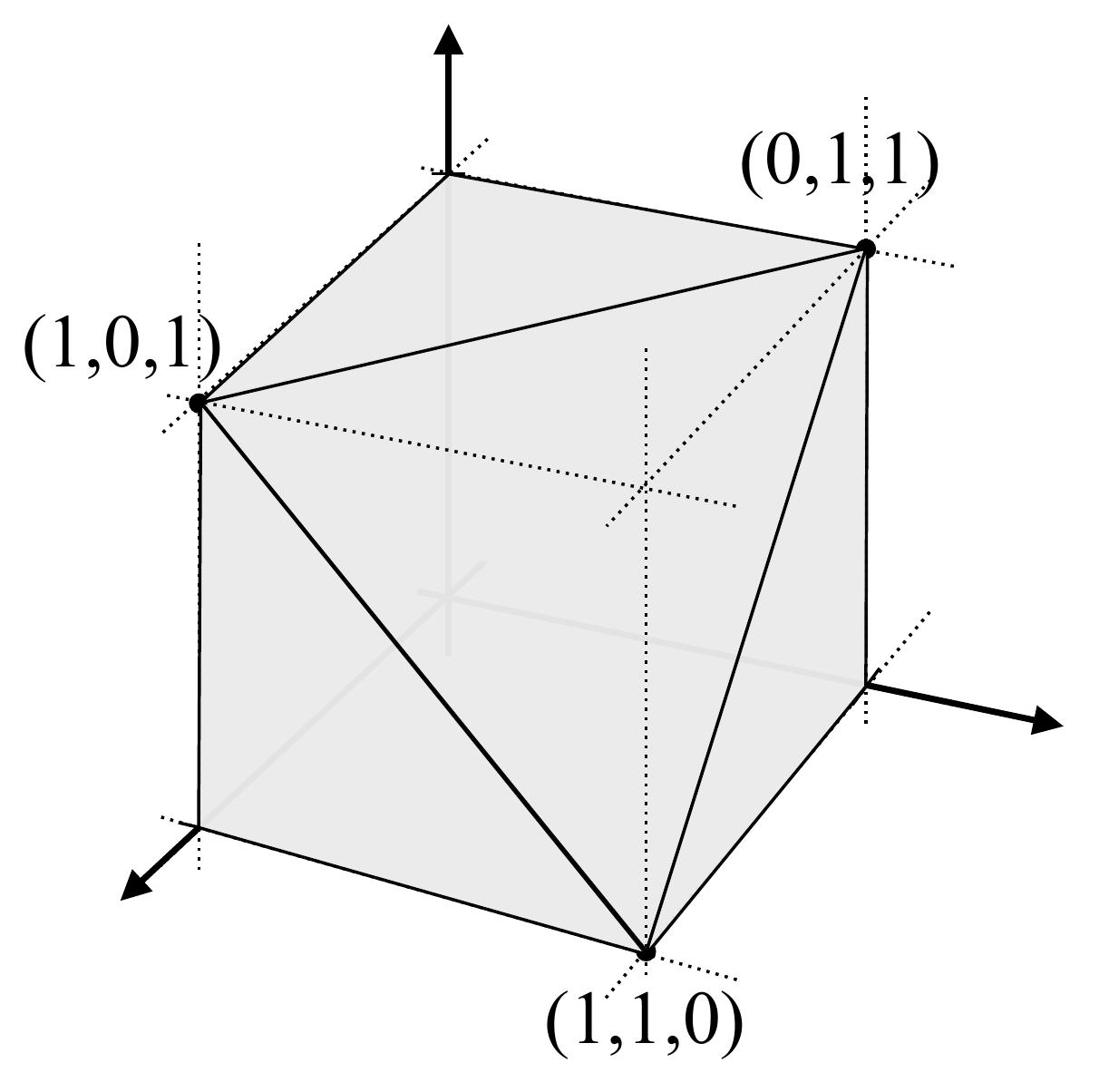}
    \end{flushright}
  \end{minipage}

  The relations $\tr$ and $\tr_\tau$ are ternary to account for the
  three calls to $f$ in the definition of $\mathtt{portest}$, and
  $\tr$ is designed so that $\tr_{D\intT}$ is as small a relation as
  possible that contains the triples
  $(\delta_1, \delta_0, 0)$ and
  $(\delta_1, 0, \delta_0)$.  Considering the three tests
  $f (\delta_1) (\delta_1) = \delta_1$,
  $f (\delta_0) (0) = \delta_0$ and
  $f (0) (\delta_0) = \delta_0$, the triple
  $(\delta_1, \delta_0, 0)$ consists of the first arguments
  to $f$ in those tests, and the triple
  $(\delta_1, 0, \delta_0)$ consists of the second arguments.
  Hence, with $f$ bound to $P$, the triple consisting of the three
  values of $f (\delta_1) (\delta_1)$,
  $f (\delta_0) (0)$ and $f (0) (\delta_0)$ respectively will
  also be contained in $\tr_{D\intT}$, by the basic lemma of logical
  relations (Proposition~\ref{prop:I:basic}).  We will then show that
  the largest probability that those values are $1$, $0$ and
  $0$ respectively is $8/27$, and this will complete the proof.

  First, let us check that $(\delta_1, \delta_0, 0)$ and
  $(\delta_1, 0, \delta_0)$ are in $\tr_{D\intT}$.  To that end, we
  simplify the expression of $\tr_{D\intT}$.  For all
  $h_1, h_2, h_3 \in \Lform \Eval {\intT}$,
  $(h_1, h_2, h_3) \in \tr_\intT^\perp$ if and only if for every
  $n \in \Eval {\intT}$, $(h_1 (n), h_2 (n), h_3 (n)) \in \tr$.  Next,
  $(a_1 \delta_0 + b_1 \delta_1, a_2 \delta_0 + b_2 \delta_1, a_3
  \delta_0 + b_3 \delta_1)$ is in $\tr_{D\intT}$ if and only if for
  all $(h_1, h_2, h_3) \in \tr_\intT^\perp$,
  $(a_1 h_1 (0) + b_1 h_1 (1), a_2 h_2 (0) + b_2 h_2 (1), a_3 h_3 (0)
  + b_3 h_3 (1)) \in \tr$.  Since $\tr$ is convex and
  downwards-closed, it suffices to check the latter when the triples
  $(h_1 (0), h_2 (0), h_3 (0))$ and $(h_1 (1), h_2 (1), h_3 (1))$ each
  range over the three points $\vec \alpha_i$, $1\leq i\leq 3$ (nine
  possibilities).  Let us write $\vec \alpha_i$ as
  $(\alpha_{i1}, \alpha_{i2}, \alpha_{i3})$.  Hence
  $(a_1 \delta_0 + b_1 \delta_1, a_2 \delta_0 + b_2 \delta_1, a_3
  \delta_0 + b_3 \delta_1)$ is in $\tr_{D\intT}$ if and only if the
  nine triples
  $(a_1 \alpha_{i1} + b_1 \alpha_{j1}, a_2 \alpha_{i2} + b_2
  \alpha_{j2}, a_3 \alpha_{i3} + b_3 \alpha_{j3})$
  ($1\leq i, j \leq 3$) are in $\tr$, namely consist of non-negative
  numbers $\leq 1$ that sum up to a value at most $2$.  Verifying that
  this holds for $(\delta_1, \delta_0, 0)$ ($a_1 \eqdef 0$,
  $b_1 \eqdef 1$, $a_2 \eqdef 1$, $b_2 \eqdef 0$,
  $a_3 \eqdef b_3 \eqdef 0$) and $(\delta_1, 0, \delta_0)$
  ($a_1 \eqdef 0$, $b_1 \eqdef 1$, $a_2 \eqdef b_2 \eqdef 0$,
  $a_3 \eqdef 1$, $b_3 \eqdef 0$) means verifying that for all $i$,
  $j$ between $1$ and $3$, $(\alpha_{j1}, \alpha_{i2}, 0)$ and
  $(\alpha_{j1}, 0, \alpha_{i3})$ are in $\tr$, which is obvious since
  those are triples of numbers equal to $0$ or to $1$.

  Using Proposition~\ref{prop:I:basic},
  $(\Eval P \allowbreak (\delta_1) \allowbreak (\delta_1), \Eval P
  (\delta_0) (0), \Eval P (0) (\delta_0))$ is also in $\tr_{D\intT}$.
  Let us write that triple as
  $(a_1 \delta_0 + b_1 \delta_1, a_2 \delta_0 + b_2 \delta_1,
  \allowbreak a_3 \delta_0 + b_3 \delta_1)$.  Then
  $\Eval {\mathtt{portest}\;P}$ is equal to
  $b_1 a_2 a_3 \cdot \delta_{0}$, as one can check.  We wish to
  maximize $b_1 a_2 a_3$ subject to the constraint
  $(a_1 \delta_0 + b_1 \delta_1, \allowbreak a_2 \delta_0 + b_2
  \delta_1, a_3 \delta_0 + b_3 \delta_1) \in \tr_{D\intT}$.  That
  constraint rewrites to the following list of twelve inequalities,
  not mentioning the constraints that say that each $a_i$ and each
  $b_i$ is non-negative:
  \begin{itemize}
  \item $a_1+b_1$, $a_2+b_2$, and $a_3+b_3$ should be at most $1$,
  \item and the nine values $a_1+b_1+a_3+b_3$, $a_1+b_1+b_2+a_3$,
    $a_1+b_2+a_3+b_3$, $a_1+b_1+a_2+b_3$, $a_1+b_1+a_2+b_2$,
    $a_1+a_2+b_2+b_3$, $b_1+a_2+a_3+b_3$, $b_1+a_2+b_2+a_3$ and
    $a_2+b_2+a_3+b_3$ should be at most $2$.
  \end{itemize}
  That is not manageable.  To help us, we have run a Monte-Carlo
  simulation: draw a large number of values at random for the
  variables $a_i$ and $b_i$ so as to verify all constraints (using
  rejection sampling), and find those that lead to the largest value
  of $b_1 a_2 a_3$.  That simulation gave us the hint that the maximal
  value of $b_1 a_2 a_3$ was indeed $8/27$, attained for
  $a_1 \eqdef 0$, $b_1 \eqdef 2/3$, $a_2 \eqdef 2/3$, $b_2 \eqdef 0$,
  $a_3 \eqdef 0$, $b_3 \eqdef 2/3$.  We now have to verify that
  formally.  Knowing which values of $a_i$ and $b_i$ maximize
  $b_1 a_2 a_3$ allows us to select which constraints are the
  important ones, and then one can simplify slightly further.

  In order to obtain a formal argument, we therefore choose to
  maximize $b_1 a_2 a_3$ with respect to the relaxed constraints that
  $a_1+b_1+a_2+b_2+a_3+b_3 \leq 2$ (an inequality implied by all the
  above constraints), all numbers being non-negative.  This will give
  us an upper bound, which may fail to be optimal (but won't).

  \begin{figure}
    \centering
    \begin{tabular}{c@{\qquad}c}
      \includegraphics[scale=0.16]{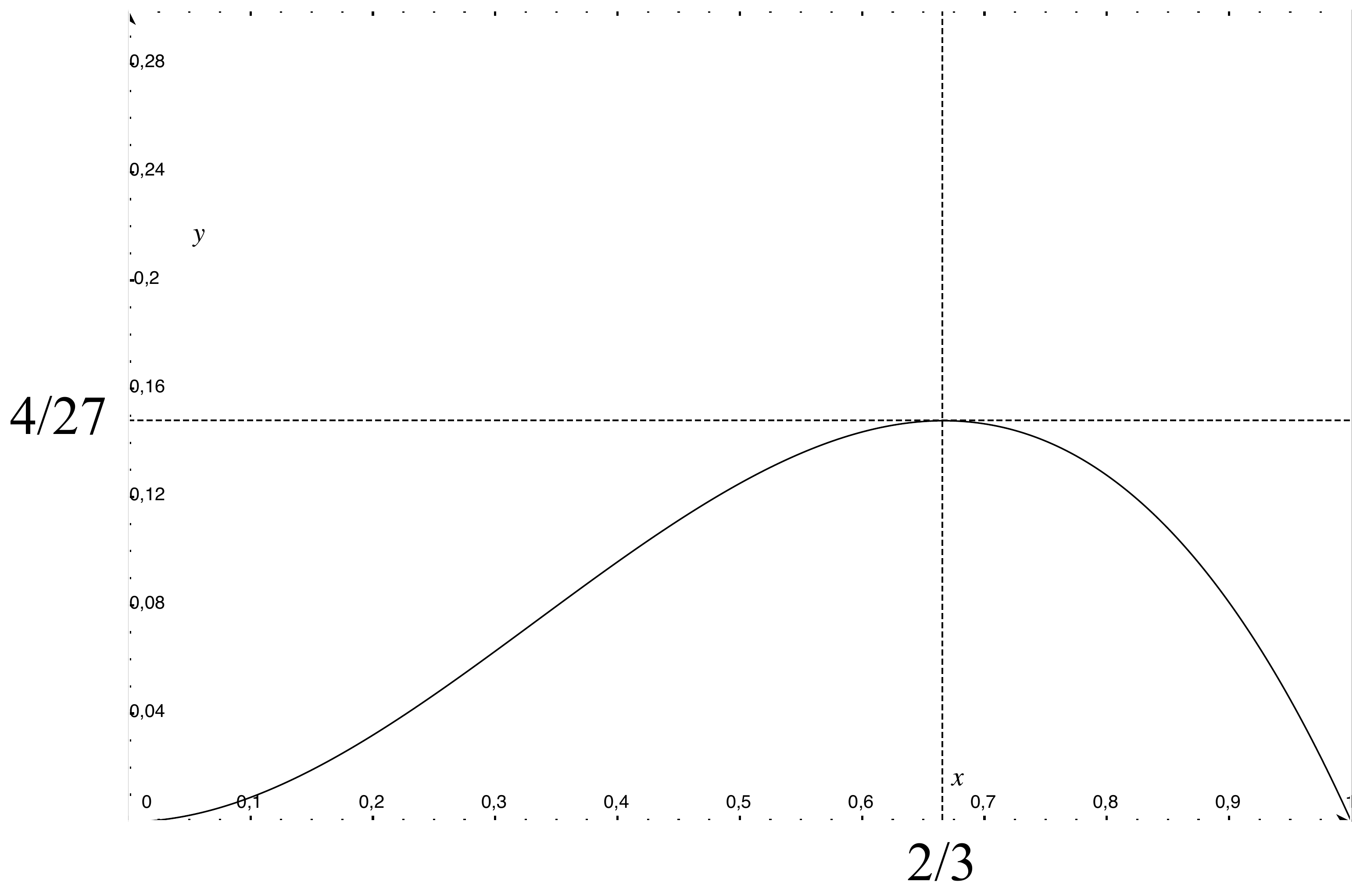}
      &
        \includegraphics[scale=0.16]{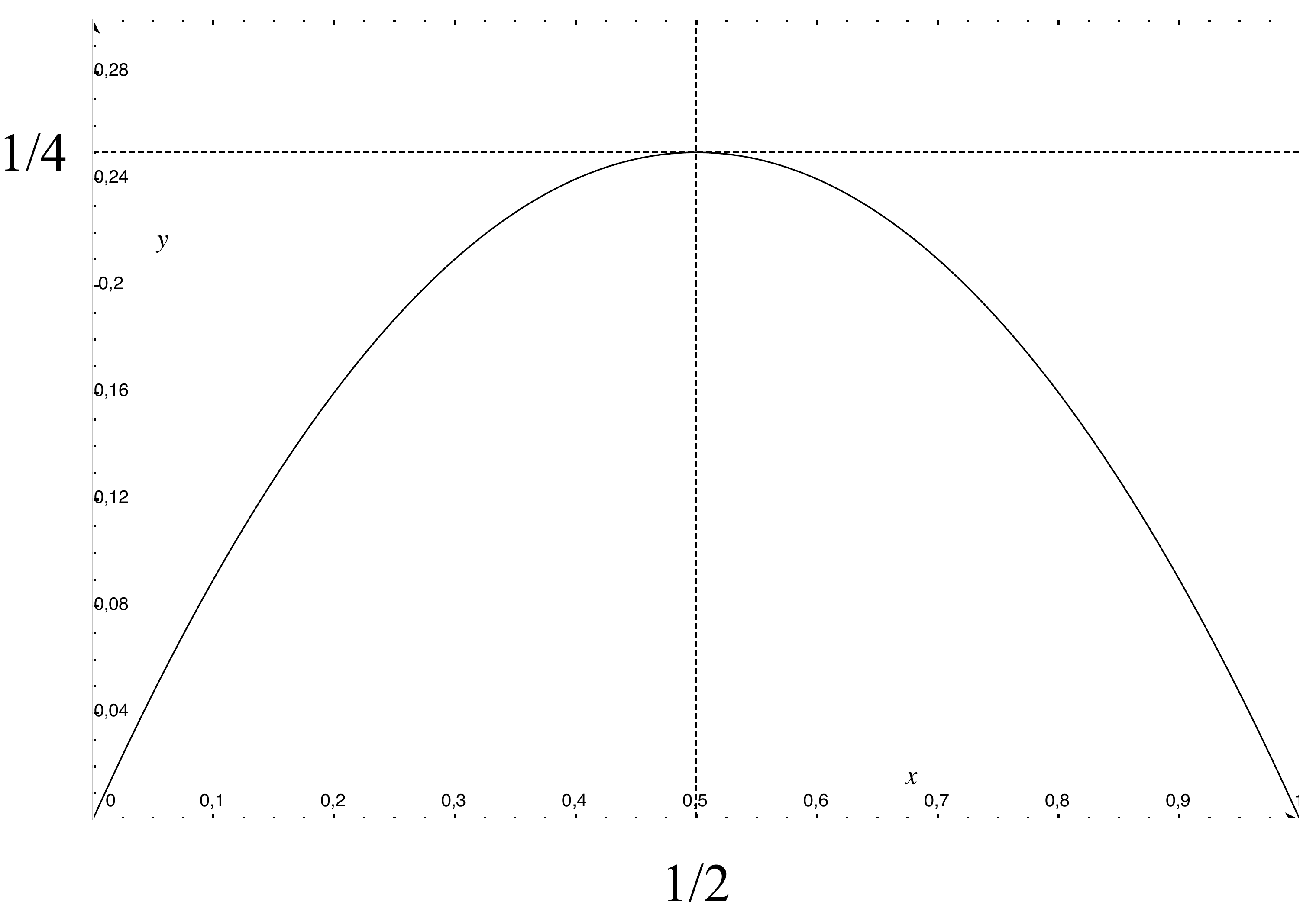} 
    \end{tabular}
    \caption{Maximizing $(1-r)r^2$ and $s(1-s)$}
    \label{fig:max}
  \end{figure}

  In order to do so, we first maximize $c_1c_2c_3$ under the
  constraints $c_1, c_2, c_3 \geq 0$ and $c_1+c_2+c_3 \leq 2$.
  Rewrite $c_1$ as $d (1-r)$, $c_2$ as $d r (1-s)$, and $c_3$ as
  $d rs$, where $d \leq 2$ and $r, s \in [0, 1]$.  (Namely, let
  $d \eqdef c_1+c_2+c_3$; if $d=0$, let $r$ and $s$ be arbitrary;
  otherwise, let $r \eqdef 1-c_1/d$; if $r=0$, then let $s$ be
  arbitrary; otherwise, let $s \eqdef c_3/(dr)$.)  The maximal value
  of $c_1c_2c_3 = d^3 (1-r)r^2 s (1-s)$ is obtained by maximizing:
  \begin{itemize}
  \item $d$ (as $2$),
  \item $(1-r)r^2$ when $r \in [0, 1]$ (value
    $4/27$ obtained at $r \eqdef 2/3$, see Figure~\ref{fig:max}, left),
  \item and $s (1-s)$ when $s \in [0, 1]$ (value $1/4$ obtained at
    $s \eqdef 1/2$, see Figure~\ref{fig:max}, right),
  \end{itemize}
  hence is equal to $2 \cdot (4/27) \cdot (1/4) = 8/27$.  It follows
  that for all $a_1, b_1, a_2, b_2, a_3, b_3 \in [0, 1]$ such that
  $a_1+b_1+a_2+b_2+a_3+b_3 \leq 2$,
  $b_1 a_2 a_3 \leq (a_1+b_1) (a_2+b_2) (a_3+b_3) \leq 8/27$, by
  taking $c_i \eqdef a_i+b_i$ for each $i$.  \qed
\end{proof}

We sum up our results as follows.  Note that $\Prob
[\mathtt{portest}\;P \vp n] = 0$, for any $P$, if $n \neq 0$.
\begin{theorem}
  \label{thm:8/27}
  For every ground PCF$_\Nature$ term
  $P \colon D\intT \to D\intT \to D\intT$, the probability
  $\Prob [\mathtt{portest}\;P \vp 0]$ that $P$ fools the parallel or
  tester never exceeds $8/27$.  That bound is attained by taking
  $P \eqdef \mathtt{pmpor}$.  \qed
\end{theorem}
\section{Conclusion and Related Work}
\label{sec:conc}

There is an extensive literature on the semantics of higher-order
functional languages, and extensions that include probabilistic choice
are now attracting attention more than ever.

Concerning denotational semantics, we should cite the following.
\emph{Probabilistic coherence spaces} provide a fully abstract
semantics for a version of PCF with probabilistic choice, as shown by
Ehrhard, Tasson, and Pagani \cite{ETP:probcoh:fa}.  \emph{Quasi-Borel
  spaces} and predomains have recently been used to give adequate
semantics to typed and untyped probabilistic programming languages,
see e.g.\ \cite{VKS:stat:dom}.  \emph{QCB spaces} form a convenient
category in which various effects, including probabilistic choice, can
be modeled \cite{Battenfeld:QCB}.  Comparatively, the domain-theoretic
semantics we are using in this paper is rather mundane, and I have
used similar models for further extensions that also include angelic
\cite{jgl-jlap14} and demonic \cite{JGL:prob:CBPV} non-deterministic
choice.  In those papers, I obtain full abstraction at the price of
adding some extra primitives, but also of considering a richer
semantics that also includes forms of non-deterministic choice.  The
latter allows us to work in categories with nice properties.  That is
not available in the context of PCF$_\Nature$, because there is no
known Cartesian-closed category of continuous dcpos that is closed
under $\Val_{\leq 1}$ \cite{JT:troublesome}.

Let me remind the reader that denotational semantics is only a tool
here: the result we have presented concerns the operational semantics,
and domain-theory is only used, through adequacy, in order to bound
$\Prob [\mathtt{portest}\;P \vp *]$.  One may wonder whether a direct
operational approach would work, but I doubt it strongly.  Eventually,
any operational approach would have to find suitable invariants, and
such invariants will be hard to distinguish from an actual
denotational semantics.

One may wonder whether such semantical proofs would be useful in the
realm of probabilistic process algebras as well.  In non-probabilistic
process algebras, syntactic reasoning is usually enough, using
bisimulations and up-to techniques.  The case of probabilistic
processes is necessarily more complex, and may benefit from such
semantical arguments.



\bibliographystyle{splncs04}
\ifspringer
\bibliography{por}
\else
\newcommand{\etalchar}[1]{$^{#1}$}

\fi

\ifspringer%
\relax%
\else%
\appendix

\section{Soundness}
\label{sec:soundness}

There is a unique way of defining a denotational semantics
$\Eval C \rho$ of contexts $C$ in such a way that
$\Eval {C [M]} \rho = \Eval C \rho (\Eval M \rho)$ for every $M$ of
the right type and every $\rho \in Env$.  For
$C \eqdef E_0 E_1 \cdots E_n$, $\Eval C \rho$ is the composition of
the maps $\Eval {E_0} \rho$, $\Eval {E_1} \rho$, \ldots,
$\Eval {E_n} \rho$, where for each elementary or initial context $E$,
$\Eval E \rho$ is defined by:
\begin{itemize}
\item for every $N \colon \sigma$, $\Eval {[\_ N]} \rho (f) \eqdef f
  (\Eval N \rho)$;
\item $\Eval {[\p \_]} \rho (n) \eqdef n-1$, $\Eval {[\suc \_]} \rho (n)
  \eqdef n+1$;
\item $\Eval {[\ifz \_ N P]} \rho (n) \eqdef \Eval N \rho$ if $n=0$,
  $\Eval P \rho$ otherwise;
\item $\Eval {[\bindkw_{\sigma,\tau} \_ N]} \rho \eqdef (\Eval N \rho)^\dagger$;
\item $\Eval {[\retkw_{\intT} \_]} \rho (n) \eqdef \eta (n) = \delta_n$;
\item $\Eval {[\_]} \rho (\nu) \eqdef \nu$;
\end{itemize}

It is standard that $\Eval M \rho$ only depends on the value of $\rho$
on the free variables of $M$ (if $\rho (x)=\rho' (x)$ for every free
variable $x$ of $M$, then $\Eval M \rho = \Eval M \rho'$), and that
for every substitution $\theta \eqdef [x_1:=N_1, \cdots, x_n:=N_n]$,
$\Eval {M \theta} \rho = \Eval M (\Eval \theta \rho)$, where
$\Eval \theta \rho$ is the environment that maps every $x_i$,
$1\leq i\leq n$, to $\Eval {N_i} \rho$ and all other variables $y$ to
$\rho (y)$.  In particular,
$\Eval {M [x_\sigma:=N]} \rho = \Eval M (\rho [x_\sigma\mapsto \Eval N
\rho])$.  Finally, $\Eval {(\lambda x_\sigma . M) N} \rho$ is equal to
$\Eval {M [x_\sigma:=N]} \rho$.  We have:
\begin{lemma}
  \label{lemma:sound}
  Let $\rho$ be an environment.
  \begin{enumerate}
  \item For every rule of the form $s \step 1 t$,
    $\Eval s \rho = \Eval t \rho$.
  \item For every context $C$ of type $D\sigma \vdash D\intT$, for all
    $M, N \colon D\sigma$,
    $\Eval {C \cdot M \oplus N} \rho = \frac 1 2 \Eval {C \cdot M}
    \rho + \frac 1 2 \Eval {C \cdot N} \rho$.
  \end{enumerate}
\end{lemma}
\begin{proof}
  1. All the cases are easily checked, except perhaps for the rule
  $C [\bindkw_{\sigma,\tau} \_ N] \cdot \retkw_\sigma M \step 1 C
  \cdot NM$.  That reduces to showing the equality
  $\Eval {\bindkw_{\sigma,\tau} (\retkw_\sigma M) N} \rho = \Eval {NM}
  \rho$.  The left-hand side is
  $(\Eval N \rho)^\dagger (\eta (\Eval M \rho))$, which is equal to
  $\Eval N \rho (\Eval M \rho)$, by (\ref{eq:dagger:eta}).  In turn,
  that is $\Eval {NM} \rho$.

  2. Let $C \eqdef E_0 E_1 \cdots E_n$.  By inspection of types, all
  the elementary contexts $E_i$, $1\leq i \leq n$, must be of the form
  $[\bindkw_{\sigma_{i+1},\sigma_i} \_ N_i]$ for some $N_i \colon
  \sigma_{i+1} \to D\sigma_i$, $E_0 = [\_]$, and $\sigma_1 = \intT$.
  
  We observe that $\Eval {E_i} \rho = (\Eval {N_i} \rho)^\dagger$ is a
  linear map.  In fact, $f^\dagger$ is linear for every
  Scott-continuous map $f \colon X \to Y$, in the following sense: for
  all $a, b \in \Rplus$ with $a+b \leq 1$, for all
  $\mu, \nu \in \Val_{\leq 1} X$,
  $f^\dagger (a\mu + b\nu) = a f^\dagger (\mu) + b f^\dagger (\nu)$.
  Indeed, for every $V \in \Open Y$,
  $f^\dagger (a \mu + b \nu) (V) = \int_{x \in X} f (x) (V)
  d(a\mu+b\nu) = a \int_{x \in X} f (x) (V) d\mu + b \int_{x \in X} f
  (x) (V) d\nu = a f^\dagger (\mu) (V) + b f^\dagger (\nu) (V)$.

  It follows that $\Eval C \rho$ is also a linear map.  Then
  $\Eval {C \cdot M \oplus N} \rho = \Eval C \rho (\frac 1 2 \Eval M
  \rho + \frac 1 2 \Eval N \rho) = \frac 1 2 \Eval C \rho (\Eval M \rho)
  + \frac 1 2 \Eval C (\Eval N \rho) = \frac 1 2 \Eval {C \cdot M}
  \rho + \frac 1 2 \Eval {C \cdot N} \rho$.  \qed
\end{proof}

\begin{proposition}[Soundness]
  \label{prop:sound}
  For every configuration $s$ of type $D\intT$, for every $n \in \Z$,
  for every environment $\rho$,
  $\Eval s \rho (\{n\}) \geq \Prob [s \vp n]$.
\end{proposition}
\begin{proof}
  It suffices to show that for every $r \in \Rplus$ such that
  $r < \Prob [s \vp n]$, $r \leq \Eval s \rho (\{n\})$.  We write
  $\Prob [s \vp V]$ as a possibly infinite sum.  Since
  $r < \Prob [s \vp n]$, there is a finite subset of the summands
  which sum to at least $r$.  In other words, there is a finite set of
  traces starting at $s$ and ending at $[\retkw_{\intT} \_] \cdot n$,
  whose weights sum up to at least $r$.  Let $N$ be some upper bound
  on the lengths of those traces.  By induction on $N$, we show that
  the sum $\Prob_{\leq N} [s \vp n]$ of all weights of traces of
  length at most $N$, starting at $s$ and ending at
  $[\retkw_{\intT} \_] \cdot n$, is less than or equal to
  $\Eval s \rho (\{n\})$, and this will prove the claim.

  If $s = [\retkw_{\intT} \_] \cdot n$, then
  $\Eval s \rho (\{n\}) = \delta_{\Eval n \rho} (\{n\})=1$.  Therefore
  $\Prob_{\leq N} [s \vp n] \leq \Eval s \rho (\{n\})$.

  From now on, we assume that $s$ is not of the form
  $[\retkw_{\intT} \_] \cdot n$.

  If $N=0$, then there is no trace of length at most $N$ starting at
  $s$ and ending at $[\retkw_{\intT} \_] \cdot n$, so
  $\Prob_{\leq N} [s \vp n] = 0 \leq \Eval s \rho (\{n\})$.

  If $N \geq 1$, then we explore three cases.

  If no rule applies to $s$, namely if $s \step a t$ for no $a$ and no
  $t$, then $\Prob_{\leq N} [s \vp n] = 0 \leq \Eval s \rho (\{n\})$.

  If $s$ if of the form $C \cdot M \oplus N$ then
  $\Prob_{\leq N-1} (C \cdot M \vp n) \leq \Eval {C \cdot M} \rho
  (\{n\})$ and
  $\Prob_{\leq N-1} (C \cdot N \vp n) \leq \Eval {C \cdot N} \rho
  (\{n\})$, by induction hypothesis.  Now
  $\Prob_{\leq N} (s \vp n) = \frac 1 2 \Prob_{\leq N-1} (C \cdot M
  \vp n) + \frac 1 2 \Prob_{\leq N-1} (C \cdot N \vp n)$, which is
  less than or equal to
  $\frac 1 2 \Eval {C \cdot M} \rho (\{n\}) +\frac 1 2 \Eval {C \cdot
    N} \rho (\{n\}) = \Eval {C \cdot M \oplus N} \rho (\{n\})$, by
  Lemma~\ref{lemma:sound}, item~2.

  In all other cases, $s \step 1 t$ for some unique configuration $t$,
  so that $\Prob_{\leq N} (s \vp n) = \Prob_{\leq N-1} (t \vp n) \leq
  \Eval t \rho (\{n\})$, by induction hypothesis.  By
  Lemma~\ref{lemma:sound}, item~1, the latter is equal to $\Eval s
  \rho (\{n\})$.  \qed
\end{proof}

\section{Adequacy}
\label{sec:adequacy}

The key to proving the converse of soundness is the design of a
suitable logical relation $\R \eqdef {(\R_\tau)}_{\tau\text{ type}}$,
where each $\R_\tau$ is a binary relation between ground terms $M$ of
type $\tau$ and elements of $\Eval \tau$.  Since $\Eval M \rho$ does
not depend on $\rho$ when $M$ is ground, we simply write $\Eval M$ in
that case.  We write $\Eval C$ similarly for ground contexts $C$.

The definition of $\R_\tau$ is by induction on $\tau$, using auxiliary
relations $\R_{D\tau}^{\perp}$ between ground contexts
$C \colon D\tau \to D\intT$ and Scott-continuous maps
$h \colon \Eval {D\tau} \to \Eval {D\intT}$:
\begin{itemize}
\item for all ground $M \colon \intT$ and $n \in \Z$, $M \R_{\intT} n$
  if and only if $[\_] \cdot M \stepstar 1 [\_] \cdot n$;
\item for all types $\sigma$, $\tau$, for all ground
  $M \colon \sigma \to \tau$ and $f \in \Eval {\sigma \to \tau}$,
  $M \R_{\sigma \to \tau} f$ if and only if for all $N \R_\sigma a$,
  $MN \R_\tau f (a)$ (we say ``for all $N \R_\sigma a$'' instead of
  ``for every ground $N \colon \sigma$ and for every
  $a \in \Eval \sigma$ such that $N \R_\sigma a$'');
\item for every type $\tau$, for all ground $M \colon D\tau$ and
  $\nu \in \Eval {D\tau}$, $M \R_{D\tau} \nu$ if and only if for every
  ground context $C \colon D\tau \vdash D\intT$, for every
  Scott-continuous map $h \colon \Eval {D\tau} \to \Eval {D\intT}$
  such that $C \R_{D\tau}^{\perp} h$, for every $n \in \Z$,
  $\Prob [C \cdot M \vp n] \geq h (\nu) (\{n\})$;
\item for every type $\tau$, for every ground context
  $C \colon D\tau \vdash D\intT$, for every Scott-continuous map
  $h \colon \Eval {D\tau} \to \Eval {D\intT}$,
  $C \R_{D\tau}^{\perp} h$ if and only if for all $P \R_\tau a$,
  for every $n \in \Z$,
  $\Prob [C \cdot \retkw_\tau P \vp n] \geq h (\eta (a)) (\{n\})$.
\end{itemize}

\begin{lemma}
  \label{lemma:R:context}
  If $C \cdot M \stepstar 1 C \cdot N$ by any sequence of rules except
  the rule
  $[\_] \cdot \retkw_{\intT} P \step 1 [\retkw_{\intT} \_] \cdot P$,
  then for every context $C'$ of the expected type,
  $C'C \cdot M \stepstar 1 C'C \cdot N$.
\end{lemma}
\begin{proof}
  It suffices to show the claim under the assumption that $C \cdot M
  \step 1 C \cdot N$ by any other rule than the one we excluded.  This
  is clear, since no rule except the one we excluded requires the
  context to have any specific shape.  \qed
\end{proof}

\begin{lemma}
  \label{lemma:step:context:factor}
  For every  context $C \colon \sigma \vdash \tau$, for
  every term $M \colon \sigma$,
  \begin{enumerate}
  \item $[\_] \cdot C [M] \stepstar 1 C \cdot M$ by using the
    exploration rules only;
  \item the run starting at $[\_] \cdot C [M]$ must start with the trace
    $[\_] \cdot C [M] \stepstar 1 C \cdot M$, followed by the run
    starting at $C \cdot M$.
  \end{enumerate}
\end{lemma}
\begin{proof}
  1 is clear.  2 is because the operational semantics is
  deterministic, in the sense that $s \step 1 t_0$ and $s \step 1 t_1$
  implies $t_0=t_1$.  \qed
\end{proof}

\begin{lemma}
  \label{lemma:R:comp:type}
  For every context $C \colon \sigma \vdash \tau$, if $\sigma$ is a
  computation type, then so is $\tau$.
\end{lemma}
\begin{proof}
  By inspection of the elementary contexts.
\end{proof}

\begin{lemma}
  \label{lemma:R:noret}
  For every configuration $s$ of type $\intT$, every trace
  $s \stepstar \alpha s'$ satisfies $\alpha=1$.  Moreover, it does not
  use the rule
  $[\_] \cdot \retkw_\beta P \step 1 [\retkw_\beta \_] \cdot P$.
\end{lemma}
\begin{proof}
  It is enough to show the claim under the assumption that
  $s \step \alpha s'$.  Let us write $s$ as $C \cdot M$, where $C$ is
  of type $\sigma \vdash \beta$ and $M \colon \sigma$.  By
  Lemma~\ref{lemma:R:comp:type}, $\sigma$ cannot be a computation
  type.  It follows that the rule that was used cannot be
  $C \cdot P \oplus Q \step {1/2} C \cdot P$ or
  $C \cdot P \oplus Q \step {1/2} C \cdot Q$, since $P \oplus Q$ has a
  computation type.  Similarly, it cannot be
  $[\_] \cdot \retkw_\beta P \step 1 [\retkw_\beta \_] \cdot P$, again
  because $\retkw_\beta P$ has a computation type.
\end{proof}

\begin{lemma}
  \label{lemma:R:reduc}
  For all terms $M \colon \tau$ and $N \colon \sigma$, for every
  context $C' \colon \sigma \vdash \tau$, for every
  $V \in \Eval \tau$, if $[\_] \cdot M \stepstar 1 C' \cdot N$ without
  using the rule
  $[\_] \cdot \retkw_{\intT} P \step 1 [\retkw_{\intT} \_] \cdot P$,
  and if $C' [N] \R_\tau V$, then $M \R_\tau V$.
\end{lemma}
\begin{proof}
  By induction on $\tau$.  If $\tau=\intT$, $C' [N] \R_\tau V$ means
  that $[\_] \cdot C' [N] \stepstar 1 [\_] \cdot V$.  By
  Lemma~\ref{lemma:step:context:factor}, item~2, our trace starting at
  $[\_] \cdot C' [N]$ and ending at $[\_] \cdot V$ must factor as
  $[\_] \cdot C' [N] \stepstar 1 C' \cdot N \stepstar 1 [\_] \cdot V$.
  Hence
  $[\_] \cdot M \stepstar 1 C' \cdot N \stepstar 1 [\_] \cdot V$,
  showing that $M \R_\tau V$.

  For types of the form $D\tau$, our task is to show that
  $M \R_{D\tau} \nu$, where $\nu$ is any subprobability valuation in
  $\Eval {D\tau}$, knowing that $C' [N] \R_\tau \nu$.  We let
  $C \colon D\tau \vdash D\intT$ be an arbitrary ground context,
  $h \colon \Eval {D\tau} \to \Eval {D\intT}$ be an arbitrary
  Scott-continuous map such that $C \R_{D\tau}^{\perp} h$, and we wish
  to show that for every $n \in \Eval {\intT}$,
  $\Prob [C \cdot M \vp n] \geq h (\nu) (\{n\})$.  By
  Lemma~\ref{lemma:R:context}, 
  $C \cdot M \stepstar 1 C C' \cdot N$.
  Since $C' [N] \R_\tau \nu$,
  $\Prob [C \cdot C' [N] \vp n] \geq h (\nu) (\{n\})$.  By
  Lemma~\ref{lemma:step:context:factor}, item~2, the run starting at
  $C \cdot C'[N]$ must factor as a trace
  $C \cdot C'[N] \stepstar 1 CC' \cdot N$ followed by a run starting
  at $CC' \cdot N$, so
  $\Prob [C \cdot C'[N] \vp n] = \Prob [CC' \cdot N \vp n]$.
  Prepending instead the trace $C \cdot M \stepstar 1 C C' \cdot N$
  (i.e., using Lemma~\ref{lemma:Prob:compute}, item~2), we see that
  $\Prob [C \cdot M \vp n] = \Prob [CC' \cdot N\vp n]$.  That is equal
  to $\Prob [C \cdot C'[N] \vp n]$, which is larger than or equal to
  $h (\nu) (\{n\})$ since $C' [N] \R_\tau \nu$ and
  $C \R_{D\tau}^{\beta\perp} h$.

  For function types $\sigma \to \tau$, we wish to show that
  $M \R_{\sigma \to \tau} f$, where $f \in \Eval {\sigma \to \tau}$,
  knowing that $C'[N] \R_{\sigma \to \tau} f$.  The latter means that
  for all $P \R_\sigma a$, $C'[N] P \R_\tau f (a)$.  For every $C$,
  there is a trace
  $C \cdot MP \step 1 C [\_ P] \cdot M \stepstar 1 C [\_ P] C' \cdot
  N$, by Lemma~\ref{lemma:R:context} with context $C [\_ P]$, and this
  trace does not use the rule
  $[\_] \cdot \retkw_\beta Q \step 1 [\retkw_\beta \_] \cdot Q$.  By
  induction hypothesis (using $[\_ P] C'$ instead of $C'$),
  $MP \R_\tau f (b)$.  Since $P$ and $b$ are arbitrary,
  $M \R_{\sigma \to \tau} f$.  \qed
\end{proof}

By taking $C' \eqdef [\_]$, we obtain the following.
\begin{corollary}
  \label{corl:R:reduc}
  Let $M, N \colon \tau$, and $V \in \Eval \tau$.  If
  $[\_] \cdot M \stepstar 1 [\_] \cdot N$ by any sequence of rules except
  $[\_] \cdot \retkw_{\intT} P \step 1 [\retkw_{\intT} \_] \cdot P$, and
  if $N \R_\tau V$ then $M \R_\tau V$.  \qed
\end{corollary}

\begin{lemma}
  \label{lemma:R:closed}
  For every ground term $M \colon \tau$, the set $M \R_\tau$, defined
  as the set of elements $a \in \Eval \tau$ such that $M \R_\tau a$,
  is Scott-closed.  If $\tau$ is a computation type, then it also
  contains the least element $\bot_\tau$ of $\Eval \tau$.
\end{lemma}
\begin{proof}
  By induction on $\tau$.  When $\tau=\intT$, this is obvious.  Let us
  consider the case of types of the form $D\tau$.  For every ground
  context $C \colon D\tau \vdash D\intT$, for every Scott-continuous
  map $h \colon \Eval {D\tau} \to \Eval {D\intT}$ such that
  $C \R_{D\tau}^{\perp} h$, for every $n \in \Eval {\intT}$, the set
  $\Gamma_{C,h,n} \eqdef \{\nu \in \Eval {D\tau} \mid h (\nu) (\{n\})
  \leq \Prob [C \cdot M \vp n]\}$ is Scott-closed: it is easily seen
  to be downwards-closed, and for every directed family
  ${(\nu_i)}_{i \in I}$ in $\Gamma_{C,h,n}$,
  $h (\dsup_{i \in I} \nu_i) (\{n\}) = \dsup_{i \in I} h (\nu_i)
  (\{n\}) \leq \Prob [C \cdot M \vp n]$, so
  $\dsup_{i \in I} \nu_i \in \Gamma_{C,h,n}$.  $M \R_{D\tau}$ is the
  intersection of all the sets $\Gamma_{C,h,n}$, hence is Scott-closed
  as well.  It also contains the least element of $\Eval {D\intT}$,
  the zero valuation, since $\Prob [C \cdot M \vp n] \geq 0$ for all
  $C$ and $n$.

  Finally, we consider function types.  Let $M \colon \sigma \to \tau$
  be ground, and let us show that $M \R_{\sigma \to \tau}$ is
  Scott-closed.  That is equal to the intersection over all
  $N \R_\sigma a$ of the sets $\Delta_{N,a}$, where
  $\Delta_{N,a} \eqdef \{f \in \Eval {\sigma \to \tau} \mid f (a) \in
  (MN \R_\tau)\}$.  $\Delta_{N,a}$ is clearly downwards-closed; for
  Scott closure, for every directed family ${(f_i)}_{i \in I}$ in
  $\Delta_{N,a}$,
  $(\dsup_{i \in I} f_i) (a) = \dsup_{i \in I} f_i (a)$ is in
  $MN \R_\tau$, because the latter is Scott-closed by induction
  hypothesis.  Taking intersections, $M \R_{\sigma \to \tau}$ is
  Scott-closed as well.

  When $\sigma \to \tau$ is a computation type, $\tau$ is one, too,
  and by induction hypothesis $MN \R_\tau \bot_\tau$ for all
  $N \R_\sigma a$.  That means that
  $MN \R_\tau \bot_{\sigma \to \tau} (a)$ for all $N \R_\sigma a$,
  hence that $M \R_{\sigma \to \tau} \bot_{\sigma \to \tau}$.
\end{proof}

\begin{corollary}
  \label{corl:R:rec}
  For every computation type $\tau$, for all $M \R_{\tau \to \tau} f$,
  $\reckw_\tau M \R_\tau \lfp_{\Eval \tau} f$.
\end{corollary}
\begin{proof}
  By the second part of Lemma~\ref{lemma:R:closed},
  $\reckw_\tau M \theta \R_\tau \bot_\tau$.

  Additionally, for every $a \in \Eval \tau$, if
  $\reckw_\tau M \R_\tau a$ then $M (\reckw_\tau M) \R_\tau f (a)$,
  since $M \R_{\tau \to \tau} f$.  Using Corollary~\ref{corl:R:reduc}
  with the step $[\_] \cdot \reckw_\tau M \step 1 M (\reckw_\tau M)$, it
  follows that $\reckw_\tau M \R_\tau f (a)$.

  Hence for every $a \in \reckw_\tau M \R_\tau$, $f (a)$ is also in
  $\reckw_\tau M \R_\tau$.  It follows that $f^n (\bot_\tau)$ is in
  $\reckw_\tau M \R_\tau$ for every $n \in \nat$.  By
  Lemma~\ref{lemma:R:closed}, $\dsup_{n \in \nat} f^n (\bot_\tau)$
  must also be in $\reckw_\tau M \R_\tau$, and that is just
  $\lfp_{\Eval \tau} (f)$.  \qed
\end{proof}

\begin{lemma}
  \label{lemma:R:ret}
  Let $\sigma$ be a type.  For all $M \R_\sigma a$,
  $\retkw_\sigma M \R_{D\sigma} \eta (a)$.
\end{lemma}
\begin{proof}
  Relying on the definition of $\R_{D\sigma}$, let $\beta$ be a basic
  type, $C \colon D\sigma \vdash D\beta$ be a ground context,
  $h \colon \Eval {D\sigma} \to \Eval {D\beta}$ be Scott-continuous,
  and assume that $C \R_{D\sigma}^{\beta\perp} h$.  By definition of
  $\R_{D\sigma}^{\beta\perp}$, and since $M \R_\sigma a$, we obtain
  $\Prob [C \cdot \retkw_\sigma M \vp V] \geq h (\eta (a))$, and that
  is what we wanted to show.
\end{proof}

\begin{lemma}
  \label{lemma:R:bind}
  Let $\sigma$, $\tau$ be types.  For all $M \R_{D\sigma} \nu$ and
  $N \R_{\sigma \to D\tau} f$, we have
  $\bindkw_{\sigma,\tau} M N \R_{D\tau} f^\dagger (\nu)$.
\end{lemma}
\begin{proof}
  We plan to use the definition of $\R_{D\tau}$, and for that we fix
  an arbitrary ground context $C \colon D\tau \vdash D\intT$, an
  arbitrary Scott-continuous map
  $h \colon \Eval {D\tau} \to \Eval {D\intT}$ such that
  $C \R_{D\tau}^{\perp} h$, and we wish to show: $(*)$ for every
  $n \in \Z$,
  $\Prob [C \cdot \bindkw_{\sigma,\tau} M N \vp n] \geq h (f^\dagger
  (\nu)) (\{n\})$.

  For all $P \R_\sigma a$, by definition of $\R_{\sigma \to D\tau}$,
  we have $NP \R_{D\tau} f (a)$.  Since $C \R_{D\tau}^{\perp} h$, and
  using the definition of $R_{D\tau}$, we obtain that
  $\Prob [C \cdot NP \vp n] \geq h (f (a)) (\{n\})$ for every
  $n \in \Z$.  We note that
  $C [\bindkw_{\sigma,\tau} \_ N] \cdot \retkw_\sigma P \step 1 C
  \cdot NP$ and we use Lemma~\ref{lemma:Prob:compute}, item~2, so
  $\Prob [C [\bindkw_{\sigma,\tau} \_ N] \cdot \retkw_\sigma P \vp n]
  \geq h (f (a)) (\{n\})$.  By (\ref{eq:dagger:eta}),
  $f = f^\dagger \circ \eta$, so
  $\Prob [C [\bindkw_{\sigma,\tau} \_ N] \cdot \retkw_\sigma P \vp n]
  \geq h (f^\dagger (\eta (a))) (\{n\})$.  Since $n$, $P$ and $a$ are
  arbitrary such that 
  $P \R_\sigma a$, we obtain that
  $C [\bindkw_{\sigma,\tau} \_ N] \R_{D\sigma}^{\perp} h \circ
  f^\dagger$, by definition of $\R_{D\sigma}^{\perp}$.

  From that and $M \R_{D\sigma} \nu$, it follows that, for every
  $n \in \Z$,
  $\Prob [C [\bindkw_{\sigma,\tau} \_ N] \cdot M \vp n] \geq h
  (f^\dagger (\nu)) (\{n\})$.  Since
  $C \cdot \bindkw_{\sigma,\tau} M N \step 1 C [\bindkw_{\sigma,\tau}
  \_ N] \cdot M$, and using Lemma~\ref{lemma:Prob:compute}, item~2, we
  obtain
  $\Prob [C \cdot \bindkw_{\sigma,\tau} M N \vp n] \geq h (f^\dagger
  (\nu)) (\{n\})$.  Since $n$, $C$ and $h$ are arbitrary such that
  $C \R_{D\tau}^{\intT\perp} h$,
  $\bindkw_{\sigma,\tau} M N \R_{D\sigma} f^\dagger (\nu)$.  \qed
\end{proof}

The crucial property of logical relations is the following \emph{basic
  lemma of logical relations}.  For a ground substitution
$\theta \eqdef [x_1:=N_1, \cdots, x_k:=N_k]$ and an environment
$\rho$, we write $\theta \R_* \rho$ to mean that for every $i$,
$1\leq i\leq k$, $N_i \R_{\tau_i} \rho (x_i)$, where $\tau_i$ is the
type of $x_i$.  The following is the \emph{basic lemma of logical
  relations} for the case at hand.
\begin{proposition}
  \label{prop:basic:lemma}
  For every PCF$_\Nature$ term $M \colon \tau$, for every ground
  substitution $\theta$ such that all the free variables of $M$ are in
  $\Dom \theta$, and for every environment $\rho$ such that
  $\theta \R_* \rho$, $M \theta \R_\tau \Eval M \rho$.
\end{proposition}
\begin{proof}
  This is by induction on the structure of $M$.  If $M = x_i$ for some
  $i$, $1\leq i\leq n$ (where $\theta = [x_1:=N_1, \cdots,
  x_n:=N_n]$), then this follows from the assumption $\theta \R_*
  \rho$.

  If $M$ is a constant $n \in \Z$, then $n \R_{\intT} n$, because
  $[\_] \cdot n \stepstar 1 [\_] \cdot n$, trivially.  If $M = \suc N$,
  then by induction hypothesis $N \theta \R_{\intT} n$, where
  $n \eqdef \Eval N \rho$.  Therefore
  $[\_] \cdot N \stepstar 1 [\_] \cdot n$.  By Lemma~\ref{lemma:R:noret},
  that trace does not use the rule
  $[\_] \cdot \retkw_\beta P \step 1 [\retkw_\beta \_] \cdot P$.  We can
  therefore apply Lemma~\ref{lemma:R:reduc} to the effect that
  $[\suc \_] \cdot N \stepstar 1 [\suc \_] \cdot n$.  Then
  $[\_] \cdot \suc M \step 1 [\suc \_] \cdot N \stepstar 1 [\suc \_]
  \cdot n \step 1 [\_] \cdot n+1 = \Eval M \rho$.  We reason similarly
  if $M = \p N$.

  In the case of terms of the form $\ifz M N P$, we must show that
  $\ifz {M \theta} {N \theta} {P \theta} \R_\tau \Eval {\ifz M N P}
  \rho$, knowing that $M \theta \R_{\intT} \Eval M \rho$,
  $N \theta \R_\tau \Eval N \rho$ and $P \theta \R_\tau \Eval P \rho$
  by induction hypothesis.  Let $n \eqdef \Eval M \rho$.  Since
  $M \theta \R_{\intT} n$, we have a trace
  $[\_] \cdot M \stepstar 1 [\_] \cdot n$, which cannot use the rule
  $[\_] \cdot \retkw_\beta Q \step 1 [\retkw_\beta \_] \cdot Q$ by
  Lemma~\ref{lemma:R:noret}.  Hence
  $[\ifz \_ {N \theta} {P \theta}] \cdot M \theta \stepstar 1 [\ifz \_
  {N \theta} {P \theta}] \cdot n$, and therefore
  $\ifz {M \theta} {N \theta} {P \theta} \stepstar 1 [\ifz \_ {N
    \theta} {P \theta}] \cdot n$ by using an additional instance of
  the leftmost exploration rule.  If $n=0$, by doing one more
  computation step, we obtain
  $\ifz {M \theta} {N \theta} {P \theta} \stepstar 1 [\_] \cdot N
  \theta$, still not using the rule
  $[\_] \cdot \retkw_\beta Q \step 1 [\retkw_\beta \_] \cdot Q$.  We now
  use Lemma~\ref{lemma:R:reduc}, and we obtain that
  $\ifz {M \theta} {N \theta} {P \theta} \R_\tau \Eval N \rho = \Eval
  M \rho$.  When $n \neq 0$, we reason similarly and we obtain that
  $\ifz {M \theta} {N \theta} {P \theta} \R_\tau \Eval P \rho = \Eval
  M \rho$.

  In the case of applications, we must show that $(MN) \theta \R_\tau
  \Eval M \rho (\Eval N \rho)$.  This follows from the definition of
  $\R_{\sigma \to \tau}$, since by induction hypothesis $M \theta
  \R_{\sigma \to \tau} \Eval M \rho$ and $N \theta \R_\sigma \Eval N \rho$.

  In the case of abstractions, we must show that
  $(\lambda x_\sigma . M) \theta \R_{\sigma \to \tau} \Eval {\lambda
    x_\sigma . M} \rho$.  We write $\theta$ as
  $[x_1:=N_1, \cdots, x_k:=N_k]$, we fix an arbitrary ground term
  $N \colon\sigma$, and a value $a \in \Eval \sigma$ such that
  $N \R_\sigma a$.  We rename $x_\sigma$ to a fresh variable if
  necessary, and we define $\theta'$ as
  $[x_1:=N_1, \cdots, x_k:=N_k, x_\sigma :=N]$, so that
  $(\lambda x_\sigma . M) \theta = \lambda x_\sigma . M \theta$ and
  $M \theta' = M \theta [x_\sigma:=N]$.  We must show that
  $(\lambda x_\sigma . M \theta) N \R_\tau \Eval M (\rho [x_\sigma
  \mapsto a])$.  By induction hypothesis,
  $M \theta' \R_\tau \Eval M (\rho [x_\sigma \mapsto a])$.  We now
  apply Corollary~\ref{corl:R:reduc}, noticing that
  $[\_] \cdot (\lambda x_\sigma . M \theta) N \step 1 [\_ N] \cdot
  \lambda x_\sigma . M \theta \step 1 [\_] \cdot M \theta
  [x_\sigma:=N] = M \theta'$.  This allows us to conclude that
  $(\lambda x_\sigma . M \theta) N \R_\tau \Eval M (\rho [x_\sigma
  \mapsto a])$, as desired.

  Let us deal with terms of the form $M \oplus N$, of type $D\tau$.
  We must show that for every ground context
  $C \colon D\tau \vdash D\intT$, for every Scott-continuous map
  $h \colon \Eval {D\tau} \to \Eval {D\intT}$ such that
  $C \R_{D\tau}^{\perp} h$, for every $n \in \Z$,
  $\Prob [C \cdot M \oplus N\vp n] \geq h (\nu) (\{n\})$.  By
  induction hypothesis, $M \theta \R_{D\tau} \Eval M \rho$, so
  $\Prob [C \cdot M\vp n] \geq h (\nu) (\{n\})$.  Similarly,
  $\Prob [C \cdot N\vp n] \geq h (\nu) (\{n\})$.  By
  Lemma~\ref{lemma:Prob:compute}, item~3,
  \begin{align*}
    \Prob [C \cdot (M \oplus N)\theta \vp n]
    & = \frac 1 2 \Prob [C \cdot M\theta \vp n]
      + \frac 1 2 \Prob [C \cdot N\theta \vp n] \\
    & \geq \frac 1 2 h (\nu) (\{n\}) + \frac 1 2 h (\nu) (\{n\}) = h
      (\nu) (\{n\}).
  \end{align*}

  The case of terms of the form $\reckw_\tau M$, $\retkw_\tau M$ and
  $\bindkw_{\sigma,\tau} M$ follow from Corollary~\ref{corl:R:rec},
  Lemma~\ref{lemma:R:ret}, and Lemma~\ref{lemma:R:bind} respectively.
  \qed
\end{proof}

\begin{lemma}
  \label{lemma:R:_}
  $[\_] \R_{D\intT}^{\perp} \identity {\Eval {D\intT}}$.
\end{lemma}
\begin{proof}
  We must show that for all $P \R_{\intT} a$, for every
  $n \in \Eval {\intT}$,
  $\Prob [[\_] \cdot \retkw_\intT P \vp n] \geq \eta (a) (\{n\})$.  By
  definition of $\R_\intT$, and since $P \R_\intT a$,
  $[\_] \cdot P \stepstar 1 [\_] \cdot a$.  By Lemma~\ref{lemma:R:noret},
  that trace does not use the rule
  $[\_] \cdot \retkw_{\intT} Q \step 1 [\retkw_{\intT} \_] \cdot Q$.  We
  can therefore use Lemma~\ref{lemma:R:context}, and we obtain
  $[\retkw_{\intT} \_] \cdot P \stepstar 1 [\retkw_{\intT} \_] \cdot
  a$.  Together with
  $[\_] \cdot \retkw_{\intT} P \step 1 [\retkw_{\intT} \_] \cdot P$, we
  obtain that
  $[\_] \cdot \retkw_{\intT} P \stepstar 1 [\retkw_{\intT} \_] \cdot a$.
  That is, $\Prob [[\_] \cdot \retkw_{\intT} P \vp n]$ is equal to $1$
  if $n=a$, $0$ otherwise.  This is precisely $\eta (a) (\{n\})$.
  \qed
\end{proof}

\begin{theorem}[Adequacy]
  \label{thm:R:adeq:1}
  For every ground term $M \colon D\intT$, for every $n \in \Z$,
  $\Eval M (\{n\}) = \Prob [M \vp n]$.
\end{theorem}
\begin{proof}
  By soundness (Proposition~\ref{prop:sound}),
  $\Eval M (\{n\}) \geq \Prob [M \vp n]$.  In the converse direction,
  we use Proposition~\ref{prop:basic:lemma} with $\theta \eqdef []$
  and we obtain $M \R_{D\intT} \Eval M$.  By Lemma~\ref{lemma:R:_},
  $[\_] \R_{D\intT}^{\perp} \identity {\Eval {D\intT}}$.  Hence, using
  the definition of $\R_{D\intT}$, for every $n \in \Z$,
  $\Prob [M \vp n] \geq \identity {\Eval {D\intT}} (\Eval M) (\{n\}) =
  \Eval M (\{n\})$.  \qed
\end{proof}

\fi%

\end{document}
